\documentclass[aps,prd,longbibliography,reprint,twocolumn,amsmath,amssymb,amsfonts,showpacs,superscriptaddress]{revtex4-2}

\usepackage{ifpdf}
\usepackage{dcolumn}
\usepackage{enumerate}
\usepackage{enumitem}
\usepackage{bm}
\usepackage{dsfont}
\usepackage{here}
\usepackage{bm}
\usepackage{float}
\floatstyle{plaintop}
\restylefloat{table}
\ifpdf

      \usepackage[pdftex]{graphicx}  
      \usepackage[pdftex]{epsfig}

\usepackage[usenames]{color}
\definecolor{navyblue}{rgb}{0.0, 0.0, 0.5}
\definecolor{ferrarired}{rgb}{1.0, 0.11, 0.0}
\definecolor{persianblue}{rgb}{0.11, 0.22, 0.73}

\usepackage[pdftex]{hyperref}
\hypersetup{
		colorlinks=true,
		citecolor=blue,%
		linkcolor=ferrarired,%
		urlcolor=persianblue,%
        filecolor=blue,
        linktoc=page
	}

\else

      \usepackage[dvips]{graphicx}  
      \usepackage[dvips]{epsfig}

      \usepackage[dvipdfm]{hyperref}
\hypersetup{
		colorlinks=true,
		citecolor=blue,%
		linkcolor=ferrarired,%
		urlcolor=persianblue,%
        filecolor=blue,
        linktoc=page
	}

\fi

\usepackage{xfrac}

\usepackage{amsmath, amsfonts}
\usepackage{nccmath}

\DeclareMathAlphabet{\mathpzc}{OT1}{pzc}{m}{it}

\usepackage{tablefootnote}

\makeatletter
\def\l@subsubsection#1#2{}
\makeatother

\begin{document}

\title{Black hole absorption cross sections: Spin and Regge poles}

\author{Mohamed \surname{Ould~El~Hadj}}\email{med.ouldelhadj@gmail.com}

\date{\today}

\begin{abstract} 
We investigate the absorption of massless scalar, electromagnetic, and gravitational fields propagating in the Schwarzschild black hole geometry. Using complex angular momentum techniques, we first derive a representation of the absorption cross section that separates it into smooth background integrals and a discrete Regge pole series. This decomposition reveals the physical mechanisms underlying black hole absorption, including classical capture, surface wave interference near the photon sphere, and subleading background effects. We then construct a refined high-frequency analytical approximation that captures both the dominant oscillations and the fine structure of the absorption spectra for scalar, electromagnetic, and gravitational fields, incorporating spin-dependent phase corrections and higher-order effects. In addition, we provide a simplified expression that generalizes the sinc approximation to describe the leading oscillations for electromagnetic and gravitational fields. Our analysis offers a unified semiclassical interpretation of black hole absorption, combining geometric optics, surface wave dynamics, and resonant phenomena encoded by the Regge pole structure.
\end{abstract}

\maketitle

\tableofcontents

\section{Introduction}
\label{sec_1}

The absorption of waves and particles by black holes plays a central role in our understanding of both classical and quantum aspects of gravity. Since the seminal works of the 1970s~\cite{Matzner:1968ufm, Misner:1972kx, Mashhoon:1973zz, Starobinsky:1973aij, Fabbri:1975sa, Ford:1975tp, Gibbons:1975kk, Page:1976df, Unruh:1976fm, Sanchez:1978sv, Sanchez:1977si}, this topic has attracted considerable attention, as it connects wave dynamics, horizon thermodynamics, and the geometric properties of black hole spacetimes. At low frequencies, a remarkable universality emerges: the absorption cross section for a minimally coupled scalar field tends to the area of the black hole’s event horizon. This behavior was first computed for the Schwarzschild black hole by Sanchez~\cite{Sanchez:1977si,Sanchez:1978sv}. A more general and explicit demonstration of this universality, valid for static and spherically symmetric black holes in arbitrary dimensions and in connection with string theory, was later provided by Das, Gibbons, and Mathur~\cite{Das:1996we}. This result, subsequently extended to rotating and higher-dimensional black holes~\cite{Maldacena:1997ih, Maldacena:1996ix, Harmark:2007jy}, has inspired numerous generalizations to fields of arbitrary spin and to alternative theories of gravity (see, e.g., Refs.~\cite{Higuchi:2001si, Kanti:2002nr, Kanti:2002ge, Harris:2003eg, Jung:2004yh, Cardoso:2005vb, Jung:2004nh, Jung:2005mr, Doran:2005vm, Grain:2005my, Crispino:2007qw, Iqbal:2008by, Dolan:2009zza, Crispino:2009ki, Crispino:2009zza, Chen:2011al, Decanini:2011xi, Decanini:2011xw, Macedo:2013afa, Benone:2014qaa, Crispino:2014eea, Leite:2016hws, Benone:2017hll, Leite:2017hkm, Leite:2018mon, Folacci:2019cmc, Folacci:2019vtt, OuldElHadj:2021fqi, OuldElHadj:2023anu, Heidari:2024bkm, Li:2025yoz}, which also provide extensive bibliographic references to earlier and related works in the field).

At high frequencies, the absorption cross section asymptotically approaches a limiting value determined by geometric optics, corresponding to the capture cross section associated with the photon sphere (see, e.g.,~\cite{Sanchez:1977si,Sanchez:1978sv,Harris:2003eg,Jung:2004yh,Jung:2005mr,Doran:2005vm,Grain:2005my,Crispino:2007qw,Dolan:2009zza,Crispino:2009ki} and references therein). However, it also exhibits characteristic oscillations around this limit, a phenomenon first explicitly computed by Sanchez~\cite{Sanchez:1977si,Sanchez:1978sv} for the Schwarzschild black hole. By performing a complete summation over partial waves, Sanchez obtained the first analytical expression for the total oscillatory behavior. While these pioneering results provided the first complete picture of the oscillatory pattern, a more general and physically unified interpretation was later offered by Decanini et al.~\cite{Decanini:2011xi,Decanini:2011xw}. Using complex angular momentum (CAM) methods, they reinterpreted these oscillations in terms of ``surface waves'' \emph{temporarily trapped} near the photon sphere, in the sense that they undergo multiple orbits along unstable circular null geodesics before escaping, and generalized the analysis to static and spherically symmetric black holes of arbitrary dimension. This CAM framework also clarified the universal character of these high-frequency oscillations, showing that their existence is intimately connected to the photon sphere's properties and is thus a generic feature of black hole spacetimes endowed with such a structure.
From a physical point of view, it is worth recalling that Regge poles characterize the dynamics of these surface, or ``Regge,'' waves near the photon sphere. Their real and imaginary parts respectively encode the dispersion relation (propagation) and damping (decay rate) of these resonant modes~\cite{Andersson:1994rm,Decanini:2009mu,Decanini:2010fz}. The CAM formalism developed by Decanini \textit{et al.} was also used to elegantly reformulate and extend the oscillatory behavior in terms of a \textit{sinc approximation}, which accurately reproduces the high-frequency fluctuations based on the Regge pole structure.

However, for higher-spin fields such as electromagnetic ($s = 1$) and gravitational ($s = 2$) perturbations, the situation becomes more complex. The onset of absorption, the structure of the effective potential, and the emergence of QNMs differ significantly from the scalar case. In particular, the oscillatory behavior of the absorption cross section for electromagnetic and gravitational fields no longer occurs around the geometric cross section, but rather around the full background contribution (combining the real- and imaginary-axis integrals) within the CAM framework, which we refer to as the \emph{background absorption cross section}.

In this work, we perform a detailed CAM-based analysis of the absorption cross section for massless fields with spins \( s = 0, 1, 2 \) in the Schwarzschild geometry. By applying the Poisson summation formula to the partial wave expansion, we construct a full CAM representation and decompose it into three physically meaningful components: a regularized real-axis background integral, an imaginary-axis contribution, and a Regge pole sum responsible for oscillatory modulations. Going beyond the scalar case treated in~\cite{Decanini:2011xi}, we highlight the distinct spin-dependent behavior of each contribution and demonstrate how their interference patterns encode fine features of the absorption spectrum across different frequency regimes. We also derive new analytic approximations incorporating spin-dependent phase corrections into the Regge pole contribution. These generalize the sinc formula and capture the dominant oscillations for electromagnetic ($s = 1$) and gravitational ($s = 2$) fields, revealing how their absorption profiles are shaped by the interplay between surface wave resonances and background absorption. Our results provide a unified and physically consistent framework for interpreting the absorption properties of Schwarzschild black holes across all massless bosonic fields.

A key result of our analysis is that the absorption cross section $\sigma_s(\omega)$ of a Schwarzschild black hole (BH) of mass $M$ immersed in a monochromatic field of spin $s$ and frequency $\omega$ is (for $2M\omega \gtrsim 1$) approximately equal to
\begin{equation}\label{Beyond_Sinc_bis}
\sigma^{\text{approx}}_s(\omega) = \sigma_{\text{geo}} \left( 1 + \frac{(1 - 6s)(1 + 6s)}{(54 M \omega)^2}  \right) + \sigma_{s,RP}(\omega)
\end{equation}
where $\sigma_{\text{geo}} = 27 \pi M^2$ is the geodesic capture cross section of the BH, and $\sigma_{s,RP}(\omega)$ is an oscillatory term with fine and hyperfine structure that arises from Regge poles which encode the effect of surface waves near the photon sphere [see Eq.~\eqref{spin_eik_Regge}].

Our article is organized as follows. In Sec.~\ref{sec_2}, we review the partial wave formalism used to compute the absorption cross section for massless bosonic fields with spins \( s = 0, 1, 2 \) propagating in the Schwarzschild geometry. We also introduce the Regge–Wheeler equation along with the appropriate boundary conditions. In Sec.~\ref{sec_3}, we construct the CAM representation of the absorption cross section. This is achieved by applying the Poisson summation formula~\cite{MorseFeshbach1953} to the partial wave expansion, leading to a decomposition into background integrals and Regge pole contributions. In other words, the CAM representation separates the cross section into an integral over the CAM plane and a discrete sum over Regge poles involving the associated residue. Sec.~\ref{sec_4} is devoted to the numerical reconstruction of the absorption cross section within the CAM formalism. After describing the computational procedure in Sec.~\ref{sec_4_1}, we present results for each spin in Sec.~\ref{sec_4_2}, comparing the CAM-based reconstruction to the exact expression obtained from the partial wave expansion. The agreement between both approaches is excellent. We also highlight some key observations concerning the crossings between the total cross section and the background contribution, which already suggest a connection with the black hole's QNM spectrum and surface wave dynamics. In Sec.~\ref{sec_5}, we derive an analytic high-frequency approximation of the absorption cross section based on the CAM decomposition. We construct a spin-dependent formula that remains accurate across a wide frequency range and validate it against numerical data. This semiclassical expression provides a reliable description of the absorption behavior not only at high frequencies but also in the intermediate and low-frequency regimes. In Sec.~\ref{sec_6}, we go beyond the standard sinc approximation by incorporating spin-dependent phase corrections into the Regge pole contribution. This yields a powerful analytic tool that captures the dominant oscillatory behavior of the absorption cross section for all spins and across frequency domains. We also examine the emergence of fine structures in the high-frequency absorption spectrum and interpret them as resulting from surface wave interference (see Sec.~\ref{sec_6_2}). Finally, Sec.~\ref{sec_7} summarizes our main results and outlines potential directions for future investigation.

Throughout this article, we adopt natural units such that \( \hbar = c = G = 1 \). We consider the exterior region of a Schwarzschild BH, described by the line element
\(ds^2 = -f(r)\, dt^2 + f(r)^{-1} dr^2 + r^2 d\theta^2 + r^2 \sin^2\theta\, d\varphi^2\), where \( f(r) = 1 - 2M/r \), and \( M \) is the mass of the BH, while $ t, r , \theta, \varphi$ are the usual Schwarzschild coordinates. In addition, we introduce the so-called \textit{tortoise coordinate} $ r_* \in ]-\infty, +\infty[$, defined by the differential relation \(dr_*/dr = 1/f(r) \), which yields the explicit expression $r_*(r) = r + 2M \ln\left(r/(2M) - 1\right)$, providing a bijection from \( r \in ]2M, +\infty[ \) to \( r_* \in ]-\infty, +\infty[ \). Finally, we assume a harmonic time dependence of the form \( e^{-i\omega t} \) for all perturbative fields.

\section{Absorption cross section for massless spin-$s$ fields}
\label{sec_2}

For a massless field of spin $s = 0$, $1$, or $2$ propagating in the Schwarzschild geometry, the total absorption cross section can be expressed as a sum over angular momentum partial waves (see e.g., Refs.~\cite{Mashhoon1973,Fabbri1975,Sanchez1978,Andersson1995}):
\begin{equation}\label{Sec_Absorption}
  \sigma_s(\omega) = \frac{\pi}{\omega^2} \sum_{\ell = s}^{+\infty} \left(2\ell+1\right) \Gamma_{\ell,s}(\omega),
\end{equation}
where the coefficients $\Gamma_{\ell,s}(\omega)$ are the \textit{greybody factors}, which quantify the absorption probability for particles with energy $\omega$ and angular momentum $\ell$. They are related to the transmission coefficients $T_{\ell,s}(\omega)$ via
\begin{equation}\label{T_coeff}
  \Gamma_{\ell,s}(\omega)= \left|T_{\ell,s}(\omega)\right|^2,
\end{equation}
where $T_{\ell,s}(\omega)$ is obtained by solving the Regge-Wheeler equation
\begin{equation}\label{Eq_RW}
  \frac{d^2}{dr_*^2}\phi_{\omega\ell s}+ \left[\omega^2-V_{\ell,s}(r)\right]\phi_{\omega\ell s} = 0.
\end{equation}
Here, the potential \( V_{\ell,s}(r) \) depends on both the spin ($s$) and angular momentum ($\ell$) of the field, and takes the form
\begin{equation}\label{Potentiel_RW}
  V_{\ell,s}(r) = f(r)\left[\frac{\left(\ell + 1/2\right)^2 - 1/4}{r^2} + (1 - s^2)\frac{2M}{r^3}\right],
\end{equation}

The functions \( \phi_{\omega\ell s}(r) \) are solutions of Eq.~\eqref{Eq_RW} that satisfy purely ingoing boundary conditions at the horizon and a combination of ingoing and outgoing waves at spatial infinity
\begin{equation}\label{phi}
\phi_{\omega\ell s}(r_*) \sim
\begin{cases}
T_{\ell,s}(\omega)\, e^{-i\omega r_*}, & r_* \to -\infty, \\
e^{-i\omega r_*} + R_{\ell,s}(\omega)\, e^{+i\omega r_*}, & r_* \to +\infty,
\end{cases}
\end{equation}
where \( R_{\ell,s}(\omega) \) denotes the reflection coefficient.

It should be noted that, in Eq.\eqref{Sec_Absorption}, \( \sigma_0(\omega) \) corresponds to the absorption cross section for a massless scalar field, while \( \sigma_1(\omega) \) and \( \sigma_2(\omega) \) describe the electromagnetic and gravitational cases, respectively. For \( s = 1 \) and \( s = 2 \), both parity sectors contribute. In the electromagnetic case, the axial and polar modes are governed by identical Regge-Wheeler equations. For gravitational perturbations, the axial sector obeys the Regge-Wheeler equation, while the polar sector is described by the Zerilli equation. However, due to an underlying isospectrality, both sectors yield identical transmission coefficients (see Refs.~\cite{Chandrasekhar:1985kt,Chandrasekhar:1975zza}), allowing the use of Eq.~\eqref{Eq_RW} for the full computation of \( \sigma_2(\omega) \).

\section{The CAM Representation of the absorption cross section based on the Poisson summation formula}
\label{sec_3}

Using the standard ``half-range'' Poisson summation formula~\cite{MorseFeshbach1953}
\begin{equation}\label{Poisson_sum}
  \sum_{\ell = 0}^{+\infty} F(\ell +1/2) = \sum_{p=-\infty}^{+\infty} (-1)^p \int_{0}^{+\infty} d\lambda\, F(\lambda)\, e^{2i\pi p\lambda},
\end{equation}
the total absorption cross section \eqref{Sec_Absorption} can be recast into a form more suitable for the application of the CAM approach. To do so, we first rewrite the sum \eqref{Sec_Absorption} as
\begin{equation}\label{Sec_Absorption_bis}
  \sigma_s(\omega) = \frac{\pi}{\omega^2} \sum_{\ell = 0}^{+\infty} \left(2\ell + 1\right) \Gamma_{\ell,s}(\omega) - \sigma_{s,\text{SM}}(\omega),
\end{equation}
where we define the subtraction term associated with the \textit{spurious modes} as
\begin{equation}\label{SP_cont}
  \sigma_{s,\text{SM}}(\omega) = \frac{\pi}{\omega^2} \sum_{\ell = 0}^{s-1} (2\ell + 1)\, \Gamma_{\ell,s}(\omega),
\end{equation}
which accounts for the artificial contributions from low angular momenta (\( \ell = 0 \) for \( s = 1 \) and \( \ell = 0, 1 \) for \( s = 2 \)) introduced to apply the Poisson summation formula. Thus, that the Poisson formula \eqref{Poisson_sum} can be applied directly to the first sum on the right-hand side, giving
\begin{fleqn}
\begin{eqnarray}\label{Appli_Poisson}
  & &\sum_{\ell = 0}^{+\infty} \left(2\ell + 1\right) \Gamma_{\ell,s}(\omega)  = \nonumber \\
  & & \hspace{30pt} \sum_{p = -\infty}^{+\infty} (-1)^p \int_{0}^{+\infty} d\lambda\, 2\lambda\, \Gamma_{\lambda - \frac{1}{2},s}(\omega)\, e^{2i\pi p\lambda}.
\end{eqnarray}
\end{fleqn}

We then separate the contributions for $p = 0$, $p > 0$, and $p < 0$, performing the change of variable $p \to -p$ in the negative-$p$ sum. This leads to
\begin{fleqn}
\begin{eqnarray}\label{Separations_diffs_Contrib}
& & \sum_{\ell = 0}^{+\infty} \left(2\ell + 1\right)\Gamma_{\ell,s}(\omega) = \int_0^{+\infty} d\lambda\, 2\lambda\, \Gamma_{\lambda - \frac{1}{2},s}(\omega) \nonumber \\
& &\hspace{35pt}+  \sum_{p = 1}^{+\infty} \int_0^{+\infty} d\lambda\, 2\lambda\, \Gamma_{\lambda - \frac{1}{2},s}(\omega)\, e^{i2\pi p \left(\lambda - \frac{1}{2}\right)} \nonumber \\
& &\hspace{35pt}+ \sum_{p = 1}^{+\infty} \int_0^{+\infty} d\lambda\, 2\lambda\, \Gamma_{\lambda - \frac{1}{2},s}(\omega)\, e^{-i2\pi p \left(\lambda - \frac{1}{2}\right)}.
\end{eqnarray}
\end{fleqn}

Substituting Eq.~\eqref{Separations_diffs_Contrib} into Eq.~\eqref{Sec_Absorption_bis}, we get the following expression for the total cross section
\begin{fleqn}
\begin{eqnarray}\label{Sec_Absorp_Poisson_Pareil}
 & & \sigma_s(\omega) = -  \sigma_{s,\text{SM}}(\omega) 
  + \frac{2\pi}{\omega^2} \int_0^{+\infty} d\lambda\, \lambda\, \Gamma_{\lambda - \frac{1}{2},s}(\omega)  \nonumber\\
 & &\hspace{35pt} + \frac{2\pi}{\omega^2} \sum_{p = 1}^{+\infty} \int_0^{+\infty} d\lambda\, \lambda\, \Gamma_{\lambda - \frac{1}{2},s}(\omega)\, e^{i2\pi p \left(\lambda - \frac{1}{2}\right)} \nonumber\\
 & &\hspace{35pt} + \frac{2\pi}{\omega^2} \sum_{p = 1}^{+\infty} \int_0^{+\infty} d\lambda\, \lambda\, \Gamma_{\lambda - \frac{1}{2},s}(\omega)\, e^{-i2\pi p \left(\lambda - \frac{1}{2}\right)}.
\end{eqnarray}
\end{fleqn}

We keep unchanged the first two terms on the right-hand side of Eq.~\eqref{Sec_Absorp_Poisson_Pareil}, i.e., the discrete sum taking into account the spurious modes (relevant when $s = 1$ or $s = 2$) and the integral along the real axis. To evaluate the third term, we apply Cauchy's residue theorem by closing the contour in the \textit{first quadrant} of the CAM plane. This involves adding an arc at infinity and extending the path along the positive imaginary axis from $+i\infty$ to $0$. Similarly, the fourth term is treated by deforming the contour to \textit{fourth quadrant}. In both cases, the deformed contours enclose the Regge poles $\lambda_{n,s}(\omega)$ with $n = 1, 2, 3, \ldots$, which correspond to the singularities of the integrands, i.e., the poles of the greybody factor $\Gamma_{\lambda-\frac{1}{2},s} (\omega)$. These poles lie in the first quadrant of the CAM plane for $\omega > 0$ and migrate to the fourth quadrant when $\omega < 0$. Assuming that the contributions from the arcs vanish at infinity, which is typically justified by exponential damping, we then obtain
\begin{align}\label{Sec_Absorption_Poisson_1}
&\sigma_s(\omega) = -  \sigma_{s,\text{SM}}(\omega) + \frac{2\pi}{\omega^2}\int_{0}^{+\infty} d\lambda \, \lambda \, \Gamma_{\ell,s}(\omega) \nonumber\\
&+\frac{2\pi}{\omega^2} \sum_{p = 1}^{+\infty} \int_{0}^{+i\infty} d\lambda \, e^{i2p\pi\left(\lambda-\frac{1}{2}\right)}\, \lambda \, \Gamma_{\lambda-\frac{1}{2},s}(\omega) \nonumber\\
&+\frac{2\pi}{\omega^2} \sum_{p = 1}^{+\infty} \int_{0}^{-i\infty} d\lambda \, e^{-i2p\pi\left(\lambda-\frac{1}{2}\right)}\, \lambda \, \Gamma_{\lambda-\frac{1}{2},s}(\omega) \nonumber\\
&+\frac{8\pi^2}{\omega^2} \operatorname{Re} \left\{ \sum_{p = 1}^{+\infty} \sum_{n = 1}^{+\infty} \, i \, e^{i2p\pi\left(\lambda_{n,s}(\omega)-\frac{1}{2}\right)} \, \lambda_{n,s}(\omega)\, \gamma_{n,s}(\omega) \right\}.
\end{align}

It is worth noting that, in establishing Eq.~\eqref{Sec_Absorption_Poisson_1}, we have followed the philosophy developed by Decanini \textit{et al.}~\cite{Decanini:2011xi} for the analytic continuation of the greybody factors into the CAM plane. Within this framework, the greybody factor is extended according to the following prescription
\begin{equation}\label{Gamma_CAM}
\Gamma_{\lambda - \frac{1}{2},s}(\omega) = T_{\lambda - \frac{1}{2},s}(\omega) \, \overline{T_{\overline{\lambda} - \frac{1}{2},s}(\omega)},
\end{equation}
where $T_{\lambda - \frac{1}{2},s}(\omega)$ denotes the analytic extension of the transmission coefficient $T_{\ell,s}(\omega)$, and the overline indicates complex conjugation.

In Eq.~\eqref{Sec_Absorption_Poisson_1}, we have also introduced the residues of the greybody factors at the Regge poles \( \lambda_{n,s}(\omega) \), which lie in the first quadrant of the complex \( \lambda \)-plane. These are given by
\begin{equation}\label{Residus}
\small
\gamma_{n,s}(\omega) = \mathop{\mathrm{Residue}}\left[T_{\lambda - \frac{1}{2},s}(\omega)\right]_{ \lambda = \lambda_{n,s}(\omega)} \overline{T_{\overline{\lambda_{n,s}(\omega)} - \frac{1}{2},s}(\omega)},
\end{equation}
and we have taken into account the contributions from the complex conjugate poles $\overline{\lambda_{n,s}(\omega)}$, located symmetrically in the fourth quadrant. Their associated residues are simply the complex conjugates $\overline{\gamma_{n,s}(\omega)}$ of those in the first quadrant.

Equation~\eqref{Sec_Absorption_Poisson_1} can now be simplified by noting that, under the analytic continuation prescription~\eqref{Gamma_CAM}, the greybody factor is even in $\lambda$, i.e., $\Gamma_{-\lambda - \frac{1}{2},s}(\omega) = \Gamma_{\lambda - \frac{1}{2},s}(\omega)$. Moreover, by applying the standard identities
\begin{fleqn}
\begin{subequations}\label{Relations}
\begin{equation}\label{Rela_1}
  \sum_{p = 1}^{+\infty} e^{i 2p \pi \left(\lambda - \frac{1}{2}\right)} = \frac{i}{2} \, \frac{e^{i\pi\left(\lambda - \frac{1}{2}\right)}}{\sin\left[\pi\left(\lambda - \frac{1}{2}\right)\right]}, \quad \text{Im} \, \lambda > 0,
\end{equation}
\begin{equation}\label{Rela_2}
  \sum_{p = 1}^{+\infty} e^{-i 2p \pi \left(\lambda - \frac{1}{2}\right)} = -\frac{i}{2} \, \frac{e^{-i\pi\left(\lambda - \frac{1}{2}\right)}}{\sin\left[\pi\left(\lambda - \frac{1}{2}\right)\right]}, \!\quad \text{Im} \, \lambda < 0,
\end{equation}
\end{subequations}
\end{fleqn}
we obtain 
\begin{eqnarray}\label{Sec_Absorption_Poisson_4}
& & \sigma_s(\omega) =  -   \sigma_{s,\text{SM}}(\omega) +  \frac{2\pi}{\omega^2}\int_{0}^{+\infty} d\lambda \, \lambda \, \Gamma_{\ell,s}(\omega) \nonumber\\
& &\hspace{8pt}-\frac{2\pi}{\omega^2}\int_{0}^{+i\infty} d\lambda \, \frac{e^{i\pi\lambda}}{\cos\left[\pi\lambda\right]}\, \lambda \, \Gamma_{\lambda-\frac{1}{2},s}(\omega)\nonumber\\
& &\hspace{8pt}-\frac{4\pi^2}{\omega^2}\text{Re}\left\{\sum_{n = 1}^{+\infty} \frac{e^{i\pi\left(\lambda_{n,s}(\omega)-\frac{1}{2}\right)}}{\sin\left[\pi\left(\lambda_{n,s}(\omega)-\frac{1}{2}\right)\right]} \, \lambda_{n,s}(\omega)\, \gamma_{n,s}(\omega)\right\}. \nonumber\\
\end{eqnarray}

We can now express the total absorption cross section in the CAM framework as a sum of distinct contributions
\begin{equation}\label{Sec_CAM_Decomp}
\sigma_s(\omega) = - \sigma_{s,\text{SM}}(\omega) + \sigma_{s,\text{B}_{\text{Re}}}(\omega) + \sigma_{s,\text{B}_{\text{Im}}}(\omega) + \sigma_{s,\text{RP}}(\omega),
\end{equation}
Here, the first term \( \sigma_{s,\text{SM}}(\omega) \) corresponds to the subtraction of spurious modes already introduced in Eq.~\eqref{SP_cont}.
The second term, \( \sigma_{s,\text{B}_{\text{Re}}}(\omega) \), corresponds to the integral over the real axis, capturing the smooth, nonresonant part of the absorption spectrum
\begin{equation}\label{Re_cont}
\sigma_{s,\text{B}_{\text{Re}}}(\omega) = \frac{2\pi}{\omega^2} \int_{0}^{+\infty} d\lambda\, \lambda\, \Gamma_{\lambda - \frac{1}{2},s}(\omega).
\end{equation}
The third term, \( \sigma_{s,\text{B}_{\text{Im}}}(\omega) \),  is an integral along the positive imaginary axis in the CAM plane. It becomes particularly relevant at low frequencies and is given by
\begin{equation}\label{Im_cont}
\sigma_{s,\text{B}_{\text{Im}}}(\omega) = -\frac{2\pi}{\omega^2} \int_{0}^{+i\infty} d\lambda\, \frac{e^{i\pi\lambda}}{\cos(\pi\lambda)}\, \lambda\, \Gamma_{\lambda - \frac{1}{2},s}(\omega).
\end{equation}
Finally, the Regge pole contribution \( \sigma_{s,\text{RP}}(\omega) \) governs the oscillatory part of the cross section and is given by
\begin{align}\label{PR_cont}
\sigma_{s,\text{RP}}(\omega) = -\frac{4\pi^2}{\omega^2} \, \text{Re} \bigg\{ &\sum_{n = 1}^{+\infty} \,\lambda_{n,s}(\omega)\, \gamma_{n,s}(\omega) \nonumber \\  
& \times \frac{e^{i\pi\left(\lambda_{n,s}(\omega)-\frac{1}{2}\right)}}{\sin\left[\pi\left(\lambda_{n,s}(\omega)-\frac{1}{2}\right)\right]} \bigg\}.
\end{align}

It should be noted that we regularize the real-axis contribution. As previously discussed, when applying the CAM formalism to the absorption cross section, the Poisson summation introduces unphysical modes for \( s = 1 \) and \( s = 2 \). These artifacts appear in the real-axis integral due to the analytic extension of the partial wave expansion. To maintain the physical consistency of the CAM decomposition, we absorb these spurious contributions into the real-axis term by defining a \emph{regularized real-axis contribution} as follows 
\begin{equation}
\widetilde{\sigma}_{s,\text{B}_{\text{Re}}}(\omega) = \sigma_{s,\text{B}_{\text{Re}}}(\omega) - \sigma_{s,\text{SM}}(\omega).
\end{equation}
For scalar fields (\( s = 0 \)), no spurious modes arise, and the regularized and unregularized real-axis contributions are identical
\begin{equation}
\widetilde{\sigma}_{s,\text{B}_{\text{Re}}}(\omega) = \sigma_{s,\text{B}_{\text{Re}}}(\omega).
\end{equation}
This prescription ensures that only the nonphysical part of the real-axis integral is removed, while the imaginary-axis  $\sigma_{s,\text{B}_{\text{Im}}}(\omega)$  and Regge pole contributions $\sigma_{s,\text{RP}}(\omega)$ remain unaltered and retain their physical interpretation. In all subsequent analysis and figures, we employ the regularized quantity \( \widetilde{\sigma}_{s,\text{B}_{\text{Re}}}(\omega) \).

\begin{figure*}[htbp]
 \includegraphics[scale=0.60]{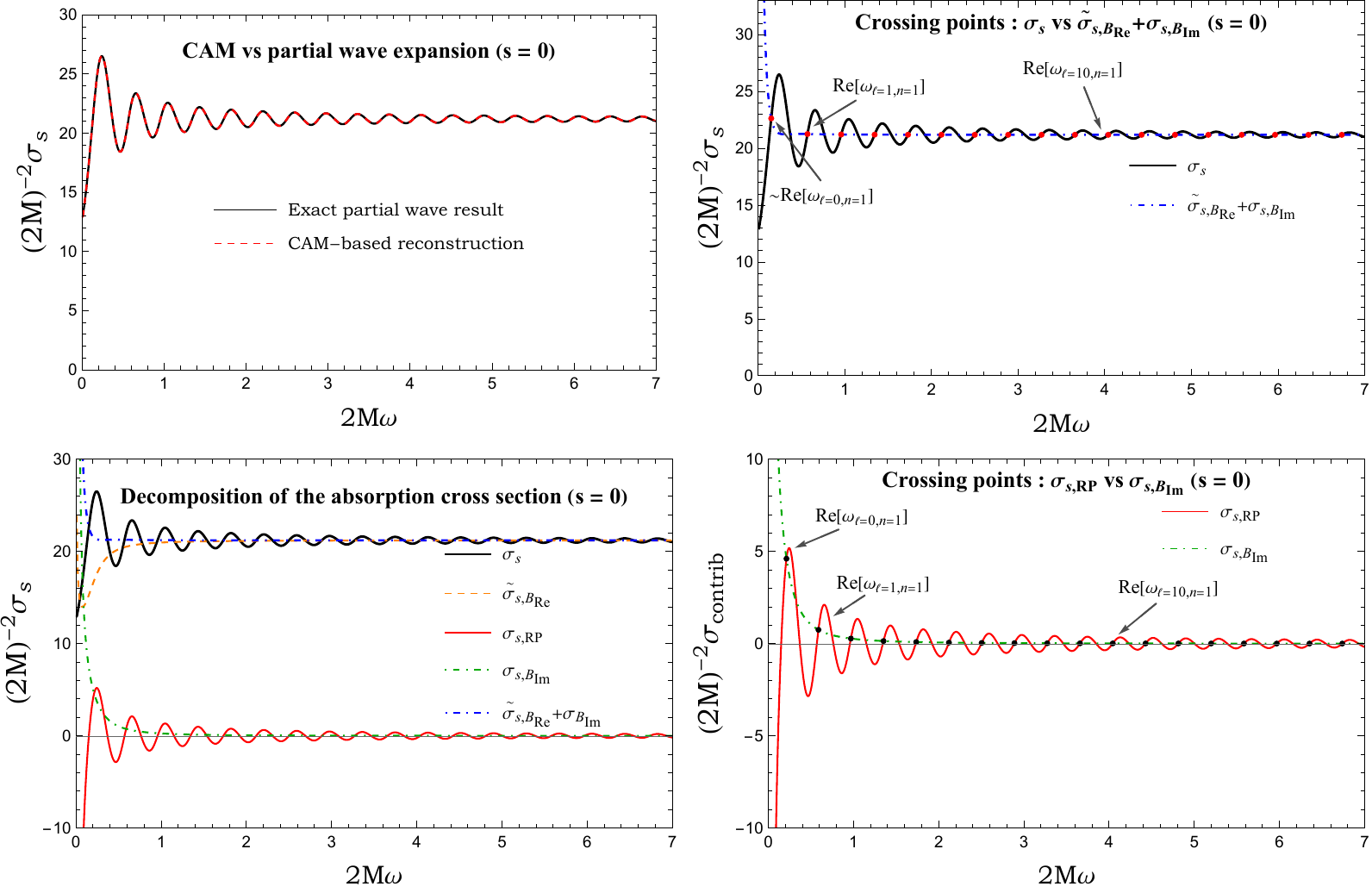}
\caption{\label{Fig:Abs_Cross_Sec_s_0} CAM-based analysis of the scalar absorption cross section ($s = 0$).
The top left panel compares the partial wave expansion with its CAM-based reconstruction. The bottom left panel shows the decomposition into Regge pole ($\sigma_{0,\text{RP}}$), imaginary background ($\sigma_{0,\text{B}_{\text{Im}}}$), and regularized real background ($\widetilde{\sigma}_{0,\text{B}_{\text{Re}}}$) contributions. In the top right, the total cross section is plotted with the background absorption cross section $\widetilde{\sigma}_{0,\text{B}_{\text{Re}}} + \sigma_{0,\text{B}_{\text{Im}}}$; their intersection points align with the real parts of fundamental QNMs for $\ell = 0,1,2,\ldots$. The bottom right panel displays $\sigma_{0,\text{RP}}$ and $\sigma_{0,\text{B}_{\text{Im}}}$, whose crossings also match $\text{Re}[\omega_{\ell,n=1}^{\text{(QNM)}}]$.}
\end{figure*}
\begin{figure*}[htbp]
 \includegraphics[scale=0.60]{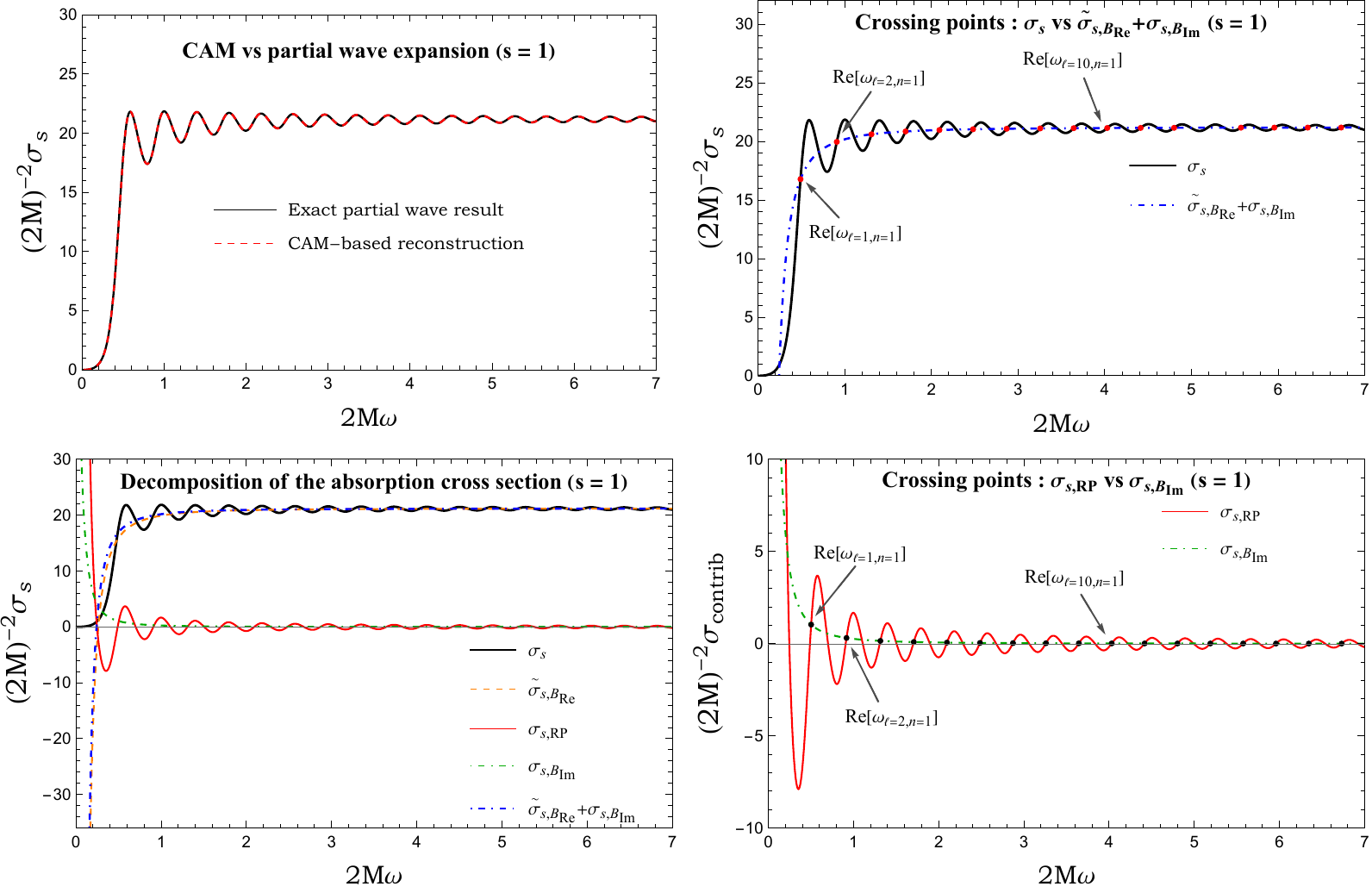}
\caption{\label{Fig:Abs_Cross_Sec_s_1} CAM-based analysis of the electromagnetic absorption cross section ($s = 1$).
The top left panel compares the partial wave expansion with its CAM-based reconstruction. The bottom left panel shows the decomposition into Regge pole ($\sigma_{1,\text{RP}}$), imaginary background ($\sigma_{1,\text{B}_{\text{Im}}}$), and regularized real background ($\widetilde{\sigma}_{1,\text{B}_{\text{Re}}}$) contributions. In the top right, the total cross section is plotted with the background absorption cross section $\widetilde{\sigma}_{1,\text{B}_{\text{Re}}} + \sigma_{1,\text{B}_{\text{Im}}}$; their intersection points align with the real parts of fundamental QNMs for $\ell = 1,2,3\ldots$. The bottom right panel displays $\sigma_{1,\text{RP}}$ and $\sigma_{1,\text{B}_{\text{Im}}}$, whose crossings also match $\text{Re}[\omega_{\ell,n=1}^{\text{(QNM)}}]$.}
\end{figure*}
\begin{figure*}[htbp]
 \includegraphics[scale=0.60]{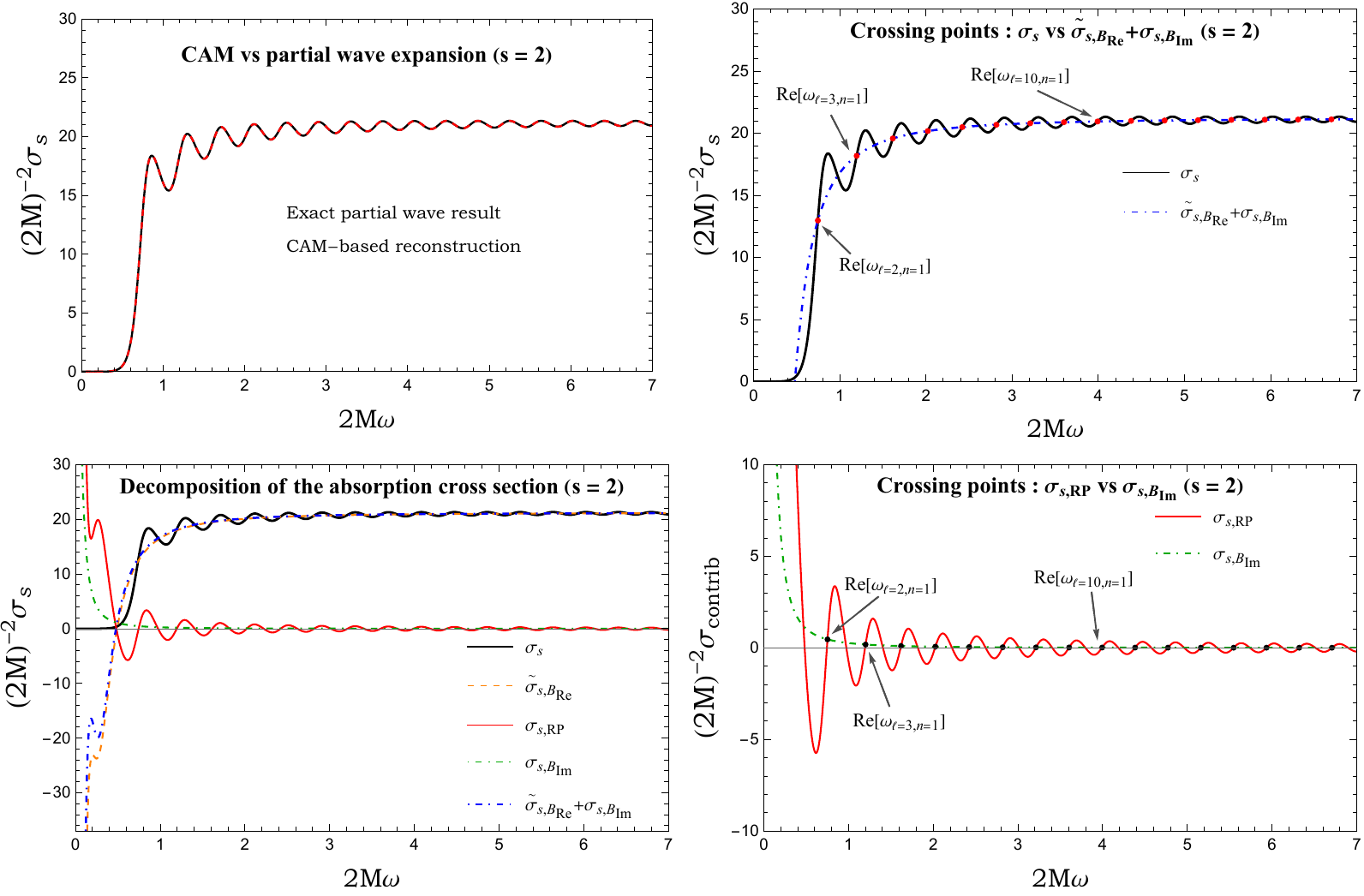}
\caption{\label{Fig:Abs_Cross_Sec_s_2} CAM-based analysis of the gravitational absorption cross section ($s = 2$)
The top left panel compares the partial wave expansion with its CAM-based reconstruction. The bottom-left panel shows the decomposition into Regge pole ($\sigma_{2,\text{RP}}$), imaginary background ($\sigma_{2,\text{B}_{\text{Im}}}$), and regularized real background ($\widetilde{\sigma}_{2,\text{B}_{\text{Re}}}$) contributions. In the top right, the total cross section is plotted with the background absorption cross section $\widetilde{\sigma}_{2,\text{B}_{\text{Re}}} + \sigma_{2,\text{B}_{\text{Im}}}$; their intersection points align with the real parts of fundamental QNMs for $\ell = 2,3,4\ldots$. The bottom right panel displays $\sigma_{2,\text{RP}}$ and $\sigma_{2,\text{B}_{\text{Im}}}$, whose crossings also match $\text{Re}[\omega_{\ell,n=1}^{\text{(QNM)}}]$.}
\end{figure*}
\begin{figure}[htbp]
 \includegraphics[scale=0.53]{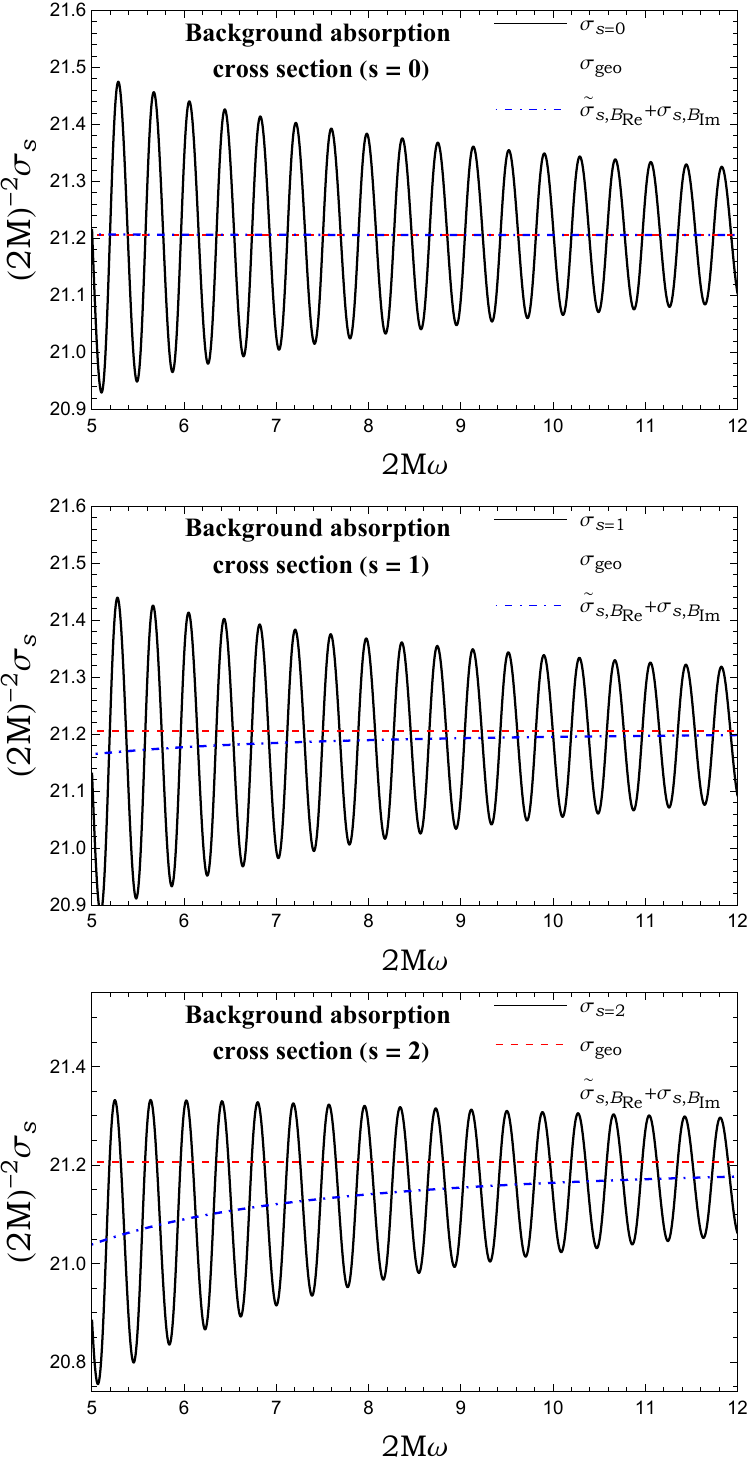}
\caption{\label{Fig:Background_Abs_s-0-1-2} High-frequency behavior of the total absorption cross section $\sigma_s(\omega)$ for spin $s$ fields in a Schwarzschild background over the range $\omega \in [5,12]$. Each panel shows the total cross section (black solid), the geometric cross section $\sigma_{\text{geo}} = 27\pi M^2$ (red dashed), and the background absorption cross section $\widetilde{\sigma}_{s,\text{B}_{\text{Re}}}(\omega) + \sigma_{s,\text{B}_{\text{Im}}}(\omega)$ (blue dot dashed). Top: scalar field ($s = 0$); middle: electromagnetic field ($s = 1$); bottom: gravitational field ($s = 2$). For $s = 0$, the cross section oscillates around $\sigma_{\text{geo}}$, while for $s = 1$ and $s = 2$, it oscillates around the  background absorption cross section.}
\end{figure}

\section{CAM Reconstruction of the Absorption Cross Section}
\label{sec_4}

\subsection{Computational methods}
\label{sec_4_1}

In order to numerically construct both the absorption cross section obtained from the partial wave expansion  Eq.~\eqref{Sec_Absorption}  and its CAM-based representation Eq.~\eqref{Sec_CAM_Decomp} [see also Eqs.~\eqref{SP_cont}--\eqref{PR_cont}], we have solved the radial equation Eqs.~\eqref{Eq_RW}--\eqref{Potentiel_RW} for both integer and complex values of the angular momentum.  For the CAM representation, the computation of the Regge pole series and the background integrals has been performed by adapting, \textit{mutatis mutandis}, the numerical method previously developed in Refs.~\cite{Folacci:2019cmc,Folacci:2019vtt} in the context of CAM-based scattering by a Schwarzschild BH, specifically for scalar and electromagnetic fields~\cite{Folacci:2019cmc}, and for gravitational perturbations~\cite{Folacci:2019vtt}. We particularly refer the reader to Sec.~IIIA of Ref.~\cite{Folacci:2019cmc} for a detailed description of the numerical procedure, and to Secs.~IIIB and IVA of Ref.~\cite{Folacci:2018sef} for the technical aspects related to the computation of Regge poles.  All numerical calculations presented in this work have been performed using \textit{Mathematica}~\cite{Mathematica13}.

\subsection{Results and interpretation}
\label{sec_4_2}

Figures~\ref{Fig:Abs_Cross_Sec_s_0}--\ref{Fig:Abs_Cross_Sec_s_2} present a detailed analysis of the absorption cross section within the CAM framework for massless fields of spin \( s = 0 \), \( 1 \), and \( 2 \). The top left panel compares the absorption cross section obtained from the partial wave expansion~\eqref{Sec_Absorption} with that reconstructed using the CAM decomposition~\eqref{Sec_CAM_Decomp} (see also Eqs.~\eqref{SP_cont} and \eqref{Re_cont}--\eqref{PR_cont}). The two curves show excellent agreement, particularly due to the inclusion of the first three Regge poles, which proves crucial for accurately capturing the low-frequency behavior as \( 2M\omega \to 0 \). The bottom left panel separately displays the contributions from the regularized real-axis background, the imaginary-axis background, their sum, and the Regge pole series. The top right panel superimposes the total cross section with the background contribution \( \widetilde{\sigma}_{s,\text{B}_\text{Re}} + \sigma_{s,\text{B}_\text{Im}} \). Remarkably, the points of intersection between the total and background cross sections coincide with the real parts of the fundamental QNM frequencies. The bottom right panel compares the Regge pole contribution to the imaginary-axis background, and once again, crossings occur at frequencies aligned with the first QNMs, reinforcing the resonant interpretation of the Regge pole series.

The CAM decomposition provides a consistent interpretation of the absorption dynamics in terms of three physically distinct contributions: (i) the Regge pole term \( \sigma_{s,\text{RP}} \), encoding the dynamics of surface waves ``trapped'' near the photon sphere~\cite{Decanini:2011xi}, (ii) the regularized real-axis background \( \widetilde{\sigma}_{s,\text{B}_\text{Re}} \), associated with classical geometric propagation, and (iii) the imaginary-axis background \( \sigma_{s,\text{B}_\text{Im}} \), which captures subleading effects such as spectral tails at low frequencies (see also Secs.~\ref{sec_5} and \ref{sec_6}). The transition between low- and high-frequency regimes is governed by the interaction between these contributions. At high frequencies, \( \widetilde{\sigma}_{s,\text{B}_\text{Re}} \) asymptotically approaches the geometric cross section (as \( \sigma_{s,\text{B}_\text{Im}} \) becomes negligible, scaling as \( 1/\omega^2 \)), while \( \sigma_{s,\text{RP}} \) generates characteristic oscillatory modulations. In contrast, at low frequencies, the relative magnitudes and signs of the three contributions depend sensitively on the spin of the field. For scalar fields (\( s=0 \)), they add constructively to reproduce the universal low-frequency limit governed by the black hole’s horizon area~\cite{Das:1996we}; whereas for electromagnetic (\( s=1 \)) and gravitational (\( s=2 \)) fields, destructive interference between the contributions leads to a strong suppression of the total absorption.

In Fig.~\ref{Fig:Background_Abs_s-0-1-2}, we observe that while the scalar absorption cross section oscillates around the classical geometric value \( \sigma_{\text{geo}} = 27\pi M^2 \) at intermediate and high frequencies, the vector and gravitational cases instead oscillate around a lower, frequency-dependent baseline, namely the sum $\widetilde{\sigma}_{s,\text{B}_{\text{Re}}} + \sigma_{s,\text{B}_{\text{Im}}}$,  which we refer to as the \emph{background absorption cross section}. This background absorption cross section corresponds to the smooth contribution in the CAM representation, which does not involve the Regge pole contributions. Physically, it represents the nonresonant, slowly varying part of the total absorption cross section, capturing the overall behavior of wave absorption by the BH in the absence of interference patterns associated with the photon sphere. At high frequencies, the background absorption cross section tends to the classical geometric-optics capture cross section $\sigma_{\text{geo}} $, in this way generalizing the concept of the geometric cross section to a frequency-dependent baseline that governs the absorption of waves with spin $s=1$ and $s=2$. Consequently, the total absorption cross section for these spins exhibits oscillations around this background, both at intermediate and high frequencies.  

Concerning the crossings between the total absorption cross section and the smooth background contribution observed in Figs.~\ref{Fig:Abs_Cross_Sec_s_0}--\ref{Fig:Abs_Cross_Sec_s_2}, we can already anticipate some important physical insights. For fields with spins \( s = 1 \) and \( s = 2 \), the first visible oscillations in the absorption cross section emerge when the frequency \( \omega \) approaches the real parts of the fundamental QNM frequencies, specifically \( \text{Re}[\omega_{\ell=1,n=1}] \sim 0.496 \) and \( \text{Re}[\omega_{\ell=2,n=1}] \sim 0.747 \), respectively. At these characteristic frequencies, the Regge pole contribution \( \sigma_{s,\text{RP}}(\omega) \) becomes locally negligible, crossing through zero. Consequently, the total absorption cross section \( \sigma_s(\omega) \) closely matches the smooth background contribution. This behavior can be qualitatively understood by noting that the oscillatory Regge pole term behaves as a sinusoidal function of the orbital phase accumulated by surface waves near the photon sphere, namely, $ \sigma_{s,\text{RP}}(\omega) \sim \sin\left(2\pi \Theta(\omega) + \delta_s(\omega)\right)$, where \( \Theta(\omega) \) encodes the orbital dynamics (see Secs.~\ref{sec_5} and \ref{sec_6}). Crossings between the total cross section and the background occur approximately when the phase satisfies the condition \( 2\pi \Theta(\omega) + \delta_s(\omega) = m\pi \) for an integer \( m \). Among these, the crossings corresponding to \( m=2\ell+1 \) align particularly well with the real parts of the fundamental QNM frequencies. Physically, these crossings signal that surface waves have accumulated specific orbital phases after circling the photon sphere, leading to constructive or destructive interference. Although they do not represent true dynamical excitations of QNMs, they nonetheless reveal the spectral imprint of the QNM structure within the absorption profile. Finally, the crossings associated with the condition $m=2\ell+1$ occur at nearly regular frequency intervals, characterized by $ \Delta\omega \sim \frac{1}{3\sqrt{3}M} = \frac{2\pi}{T_\text{orb}}$, where \( T_\text{orb} \) is the orbital period of massless particles on the photon sphere. This spacing reflects the underlying semiclassical structure of the BH geometry, and highlights the intimate connection between the oscillatory pattern of the absorption cross section and the unstable photon orbits.

\section{High-Frequency CAM-based formula for the absorption cross section}
\label{sec_5}

Building upon the CAM formalism, this section presents an analytic approximation of the total absorption cross section in the high-energy regime. The formula encapsulates both the smooth background absorption cross section, which generalizes the geometric limit beyond the scalar case, and the leading-order oscillatory corrections arising from the Regge pole structure. We begin from the decomposition in Eq.~\eqref{Sec_CAM_Decomp}, and construct asymptotic approximations for each of the terms contributing to the total absorption cross section.

\subsection{CAM-based approximation: Theoretical construction}
\label{sec_5_2}

To construct the explicit contribution of the spurious modes, we analyze the corresponding low multipoles that appear in the CAM decomposition but do not represent physical degrees of freedom. For the electromagnetic monopole mode $(s = 1, \ell = 0)$, the effective potential in the Regge-Wheeler equation vanishes identically, and the radial equation reduces to that of a free wave. Imposing ingoing boundary conditions at the horizon yields a purely ingoing solution throughout the entire domain, implying full transmission with no reflection. As a result, the associated greybody factor is exactly
\begin{equation}
\Gamma_{1,0}(\omega) = 1.
\end{equation}

For the gravitational sector $(s = 2)$, the spurious contributions arise from the $\ell = 0$ and $\ell = 1$ modes, which are governed by nonzero effective potentials. While partial reflection occurs at low frequencies, in the high-frequency regime $\omega \gg 1/(2M)$, the potential barrier becomes negligible in comparison to the wave energy, and the system enters the semiclassical regime. In this limit, the transmission is nearly perfect, and the greybody factors tend to unity
\begin{equation}
\lim_{\omega \to \infty} \Gamma_{2,\ell}(\omega) = 1, \quad \text{for } \ell = 0, 1.
\end{equation}
Accordingly, the spurious mode contribution to the absorption cross section simplifies to
\begin{equation}\label{Asump_SP_cont}
\sigma_{s,\text{SP}}(\omega) = \frac{\pi}{\omega^2} \sum_{\ell = 0}^{s-1} (2\ell + 1)\, \Gamma_{\ell,s}(\omega) \approx \frac{\pi}{\omega^2} s^2.
\end{equation}

To construct the real-axis background contribution in the eikonal (high-frequency) regime, we make use of the WKB approximation. In this context, it has been shown that the greybody factors $\Gamma_{\ell,s}(\omega)$ can be approximated at leading order by~\cite{Iyer1987,Iyer1987bis}
\begin{equation}\label{WKB_Greybody}
\Gamma_{\ell,s}^{\text{(WKB)}}(\omega) = \frac{1}{1 + \exp\left[ -\frac{2\pi (\omega^2 - V_0(\ell))}{\sqrt{-2 V_0^{\prime\prime}(\ell)}} \right]},
\end{equation}
which describes the transmission probability near the top of the potential barrier, i.e., in the region where $\omega^2 \approx V_0(\ell)$. Here, \( V_0(\ell) \) and \( V_0^{\prime\prime}(\ell) \) denote, respectively, the height and curvature of the effective potential evaluated at its maximum in the tortoise coordinate. In the large-\( \ell \) limit, they admit the following asymptotic expansions
\begin{fleqn}
\begin{subequations}
\begin{align}
&V_0(\ell) \equiv V_{\ell,s}(r_*) \bigg|_{{r_*} =  r_*^{\text{max}}} = \nonumber\\ 
& \hspace{15pt} \frac{(l + 1/2)^2}{27 M^2} +\frac{8(1-s^2)-3}{81} + \underset{\ell \to \infty}{\mathcal{O}}(1/\ell^2), 
\end{align}
\begin{align}
& V_0^{\prime\prime} (\ell)  \equiv  \frac{d^2}{dr_*^2}  V_{\ell,s}(r_*)\bigg|_{{r_*}=r_*^{\text{max}}}  =  \nonumber \\
&-\frac{2(l + 1/2)^2}{(27 M^2)^2} -\frac{8(32(1-s^2)-9)}{6561} + \underset{\ell \to \infty}{\mathcal{O}}(1/\ell^2).
\end{align}
\end{subequations}
\end{fleqn}

To facilitate the integration of the real-axis contribution \( \sigma_{s,\text{B}_\text{Re}} \), we approximate the WKB greybody factors~\eqref{WKB_Greybody} by a smooth profile based on the error function (Erf),
\begin{equation}
\frac{1}{1 + e^{f(x)}} \approx \frac{1}{2} \left[ 1 + \mathrm{Erf}\left( -\frac{f(x)}{\sqrt{2\pi}} \right) \right],
\end{equation}
which captures the transition behavior of the greybody factor near the top of the potential barrier. This leads to a analytic model for the greybody factor, extended to CAM, of the form
\begin{equation}
\Gamma_{\lambda-\frac{1}{2}, s}^{\text{(Erf)}}(\omega, \lambda) = \frac{1}{2} \left\{ 1 + \mathrm{Erf} \left[ -\kappa_s \left( \lambda - \frac{27 M^2 \omega^2}{\lambda} \right) \right] \right\},
\end{equation}
where \( \kappa_s \) is a slope parameter encoding the curvature of the potential barrier.

In principle, \( \kappa_s \) can be estimated from the WKB expression~\eqref{WKB_Greybody} using asymptotic expansions of the potential. However, since the WKB result is strictly valid only in the high-frequency limit (\( \omega \gg 1/(2M) \)), it does not fully capture the transmission spectrum at intermediate frequencies. Therefore, we treat \( \kappa_s \) as an effective fitting parameter, calibrated numerically to optimize agreement with the exact cross section across a broad frequency range. This approach yields a semianalytic fit that retains the correct high-frequency asymptotics while improving accuracy in the transition region. For electromagnetic (\( s = 1 \)) and gravitational (\( s = 2 \)) fields, we propose the following slope parameters
\begin{equation}
\kappa_s = 
\begin{cases}
\displaystyle \frac{3\sqrt{3}}{8}, & \text{for } s = 1, \\[6pt]
\displaystyle \frac{3\sqrt{3}}{4\sqrt{17}}, & \text{for } s = 2,
\end{cases}
\end{equation}
which lead to the following expression for the regularized real-axis contribution:
\begin{align}\label{Asymp_Re_Cont}
  \sigma_{s,\text{B}_\text{Re}}(\omega)  &= \frac{2\pi}{\omega^2} \int_0^\infty \lambda\, \Gamma_s^{\text{(Erf)}}(\omega, \lambda)\, d\lambda \nonumber\\
 & \approx 27\pi M^2 \left[1 - \frac{2(1 + 3s)(1 - 3s)}{(27 M \omega)^2} \right].
\end{align}
This result reproduces the geometric cross section \( \sigma_{\text{geo}} = 27\pi M^2 \) in the eikonal limit and incorporates a subleading correction that improves accuracy at intermediate frequencies.

We now consider the third contribution, corresponding to the background integral along the imaginary axis \( \sigma_{s,\text{B}_\text{Im}}(\omega) \). The greybody factors \( \Gamma_{\lambda - \frac{1}{2}, s}(\omega) \), when evaluated along the positive imaginary axis in the complex $\lambda$-plane, is a smooth and well-behaved function. It has no poles in this region. In the high-frequency regime, it is a good approximation to assume that \( \Gamma_{\lambda - \frac{1}{2}, s}(\omega) \to 1 \) along the entire integration contour~\cite{Decanini:2011xi}. Under this assumption, the background integral simplifies and yields the following asymptotic result
\begin{align}\label{Asymp_Im_cont}
\sigma_{s,\text{B}_\text{Im}}(\omega) &= -\frac{2\pi}{\omega^2} \int_{0}^{+i\infty} d\lambda\, \frac{e^{i\pi\lambda}}{\cos(\pi\lambda)}\, \lambda\, \Gamma_{\lambda - \frac{1}{2},s}(\omega) \nonumber\\ 
& \approx \frac{\pi}{12 \omega^2}
\end{align}

We finally consider the fourth contribution, associated with the Regge pole series, which encodes the leading oscillatory behavior of the absorption cross section. In the high-frequency regime, the dominant contribution comes from the first Regge pole, leading to the following expression
\begin{align}\label{Asymp_PR_cont}
\sigma_{s,\text{RP}}(\omega) = -\frac{4\pi^2}{\omega^2} \, \text{Re} \bigg\{ &\,\lambda_{1,s}(\omega)\, \gamma_{1,s}(\omega) \nonumber \\  
& \times \frac{e^{i\pi\left(\lambda_{1,s}(\omega)-\frac{1}{2}\right)}}{\sin\left[\pi\left(\lambda_{1,s}(\omega)-\frac{1}{2}\right)\right]} \bigg\}
\end{align}
with Regge pole \( \lambda_{n,s}(\omega) \) and the associated residues \( \gamma_{n,s}(\omega) \) can be computed using the WKB method and are given by~\cite{Decanini:2002ha,Decanini:2009mu,Decanini:2011xw}
\begin{widetext}
\begin{equation}\label{PR_WKB}
\small
\begin{split}
  & \lambda_{n,s}(\omega) =  3\sqrt{3} M \omega + i\alpha+ \frac{1}{M \omega}\left[\frac{115-144\beta+ 60\alpha^2}{1296\sqrt{3}}\right] 
   - \frac{i\alpha}{\left(M\omega\right)^2}\left[\frac{5555 - 6912\beta + 1220 \alpha^2}{419904}\right] \\
   &+\frac{1}{\left(M\omega\right)^3 }\left[\frac{-2079661 + 4966272 \beta - 2985984 \beta^2 + 24(-807595 + 1000512 \beta) \alpha^2 - 2357520 \alpha^4}{3265173504 \sqrt{3}}\right] \\
   &+\frac{i \alpha }{\left(M\omega\right)^4}\left[\frac{593617841 - 1425185280 \beta + 859963392 \beta^2 + 120 \alpha^2 (15598279 - 19263744 \beta) + 144920784 \alpha^4}{2115832430592}\right]\\
   &+ \underset{M\omega \to \infty}{\mathcal{O}}(\frac{1}{(M\omega)^5})
  \end{split}
\end{equation}
and
\begin{equation}\label{Residues_WKB}
   \begin{split}
   &\gamma_{n,s}(\omega) = -\frac{1}{2\pi} +\frac{i \alpha}{M\omega}\left[\frac{5}{108 \sqrt{3} \pi}\right] 
   +\frac{1}{\left(M\omega\right)^2} \left[\frac{5555 + 3660 \alpha^2 - 6912 \beta}{839808  \pi}\right] \\
   &-\frac{i \alpha }{\left(M\omega\right)^3} \left[\frac{807595 + 196460 \alpha^2 - 1000512 \beta }{136048896 \sqrt{3} \pi}\right]\\
   &-\frac{1}{\left(M\omega\right)^4} \left[\frac{724603920 \alpha ^4+360 \alpha ^2 (15598279-19263744 \beta )+124416 \beta  (6912 \beta -11455)+593617841}{4231664861184 \pi }\right]\\
   &+ \underset{M\omega \to \infty}{\mathcal{O}}(\frac{1}{(M\omega)^5})
   \end{split}
\end{equation}
\end{widetext}
where
\begin{equation}\label{alpha}
  \alpha = n - \frac{1}{2}, \qquad \beta = 1 - s^2.
\end{equation}

Taking into account all contributions discussed above, namely the spurious modes Eq.~\eqref{Asump_SP_cont}, the real-axis background Eq.~\eqref{Asymp_Re_Cont}, the imaginary-axis background Eq.~\eqref{Asymp_Im_cont}, and the Regge pole series Eq.~\eqref{Asymp_PR_cont}, we arrive at the following ``high-frequency'' expression for the total absorption cross section
\begin{widetext}
\begin{equation}\label{Beyond_Sinc}
    \sigma_s^{\text{approx}}(\omega) = -\frac{\pi}{\omega^2}s^2 + \frac{\pi}{12\,\omega^2}  + 27\pi M^2 \left[1 - \frac{2(1 + 3s)(1 - 3s)}{(27 M \omega)^2}\right] + \sigma_{s,\text{RP}}(\omega).
\end{equation}
\end{widetext}
By appropriately combining these contributions, one obtains the more compact expression given in Eq.~\eqref{Beyond_Sinc_bis}.

\subsection{Results of CAM-based approximation}
\label{sec_5_3}

\begin{figure}[htbp]
 \includegraphics[scale=0.53]{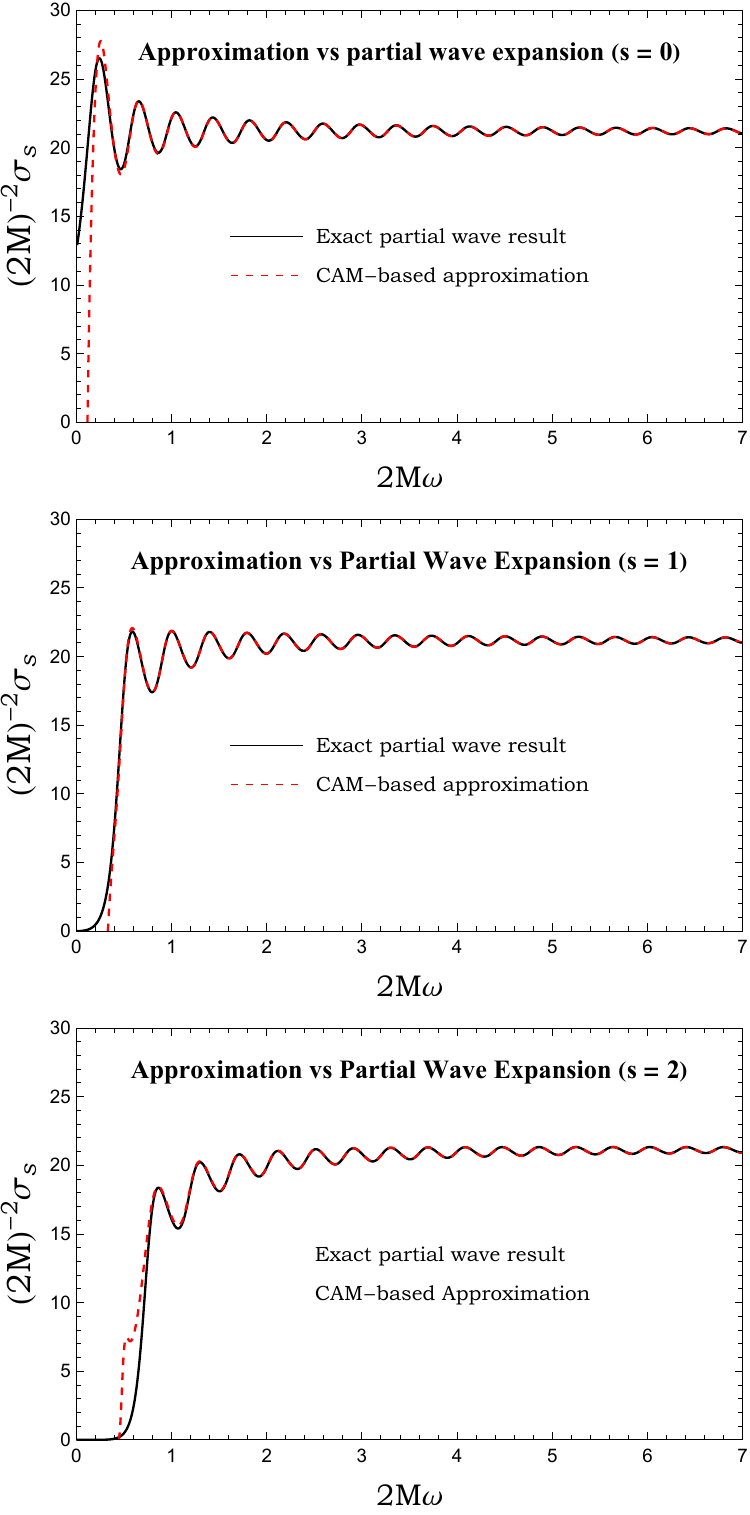}
\caption{\label{Fig:Approx_Abs_s-0-1-2} Comparison between the total absorption cross section obtained from the partial wave expansion \eqref{Sec_Absorption} and the analytic CAM-based approximation for different spin values \eqref{Beyond_Sinc}. Top: scalar field ($s = 0$); middle: electromagnetic field ($s = 1$); bottom: gravitational field ($s = 2$).}
\end{figure}

In Fig.~\ref{Fig:Approx_Abs_s-0-1-2}, we present the results obtained from the CAM-based approximation given in Eq.~\eqref{Beyond_Sinc}. In particular, the Regge pole contribution was computed by inserting the analytic expressions of the Regge poles \( \lambda_{n=1,s}(\omega) \) and their associated residues \( \gamma_{n=1,s}(\omega) \) into Eq.~\eqref{Asymp_PR_cont}, as provided by Eqs.~\eqref{PR_WKB} and \eqref{Residues_WKB}. For the scalar case ($s = 0$), since no refined approximation of the greybody factor beyond the geometric cross section was used, only the leading-order term in the real-axis background contribution was retained. This corresponds to the geometric limit. Accordingly, the Regge pole expansion was included up to and including order \( \mathcal{O}(1/(M\omega)) \), while the imaginary-axis contribution was neglected, as it enters at order \( \mathcal{O}(1/(M\omega)^2) \). In the electromagnetic case ($s = 1$), all four contributions were taken into account: the spurious modes, the real-axis background using an error-function approximation of the greybody factor, the imaginary-axis background, and the Regge pole series, which was included up to and including order \( \mathcal{O}(1/(M\omega)^3) \) for both pole positions and residues. Similarly, for the gravitational case ($s = 2$), the Regge pole expansion was included up to and including order \( \mathcal{O}(1/(M\omega)^4) \), in order to ensure excellent agreement with the full partial wave expansion across a broad frequency range, including intermediate and low frequencies.

\begin{figure}[htbp]
 \includegraphics[scale=0.53]{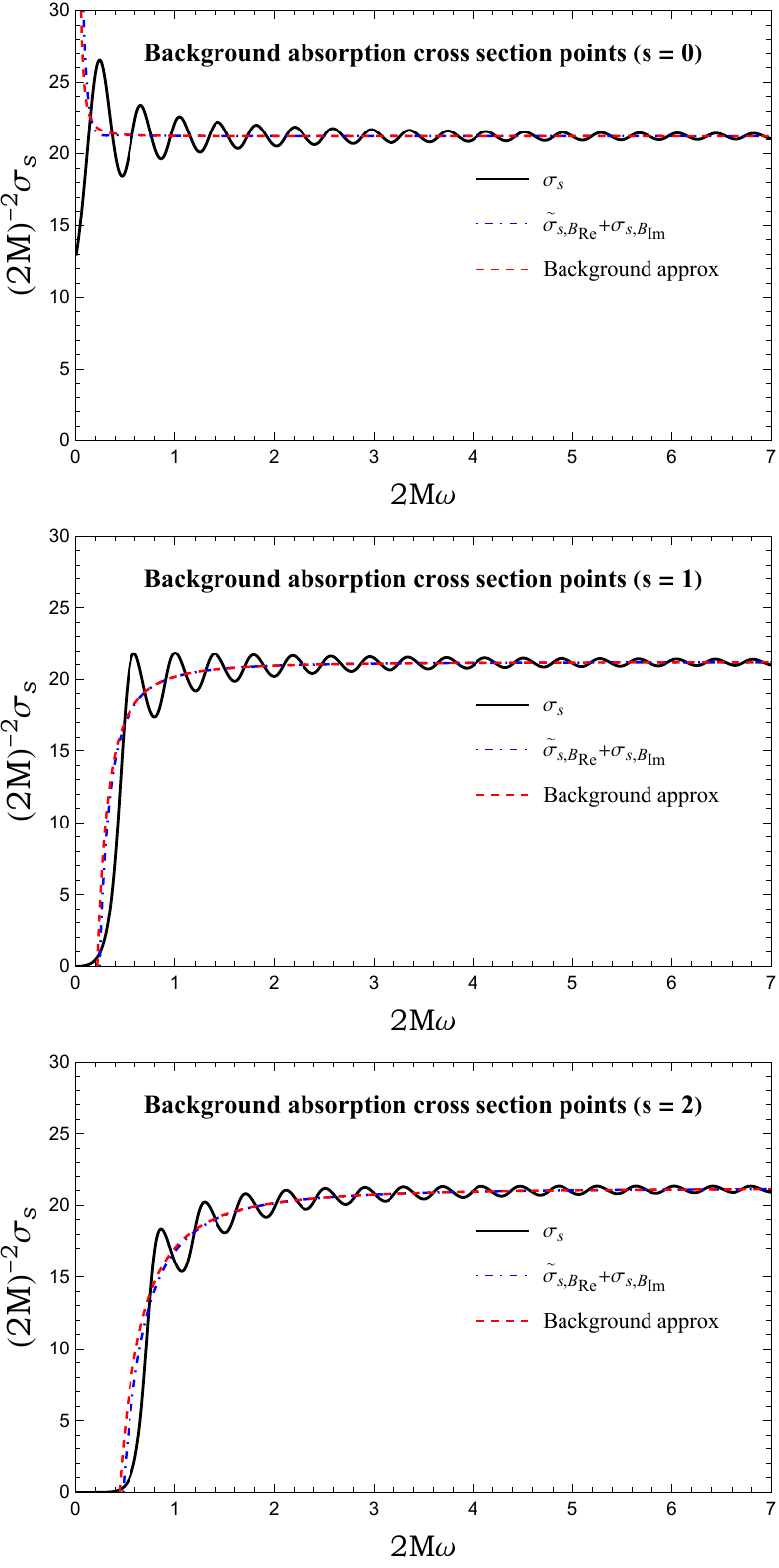}
\caption{\label{Fig:Approx_Background_Abs_s-0-1-2} Comparison between the numerical background contribution $\widetilde{\sigma}_{\text{B}_\text{Re}} + \sigma_{\text{B}_\text{Im}}$ (blue dot dashed line), and its analytic approximation (red dashed line). Top: scalar field ($s = 0$), middle: electromagnetic field ($s = 1$),  bottom: gravitational field ($s = 2$).}
\end{figure}

In Fig.\ref{Fig:Approx_Background_Abs_s-0-1-2}, we test the quality of the analytic background approximation by comparing it to the exact numerical evaluation of \( \widetilde{\sigma}_{s,\text{B}_\text{Re}}(\omega) + \sigma_{s,\text{B}_\text{Im}}(\omega) \)  for different spin fields. For the scalar case ($s = 0$), no dedicated error-function model was constructed for the greybody factor \(\Gamma_{\ell,s}(\omega)\), and the analytic approximation of the real-axis integral was restricted to the leading geometric contribution. Nevertheless, by employing the real-axis approximation given in Eq.\eqref{Asymp_Re_Cont} together with the imaginary-axis contribution in Eq.~\eqref{Asymp_Im_cont}, we find that this combined functional form offers a remarkably good description of the scalar background as well, despite the absence of a specific theoretical justification at this stage. For the electromagnetic and gravitational cases ($s = 1$ and $s = 2$), the analytic background approximation, including both the regularized real-axis and imaginary-axis contributions, shows excellent agreement with the numerical results across the full frequency range. This validates the accuracy and robustness of the proposed error-function-based modeling for smooth background contributions within the CAM formalism. It is therefore important to emphasize that the final high-frequency formula \eqref{Beyond_Sinc} (cf. Eq.~\eqref{Beyond_Sinc_bis}) remains valid for all three spins considered ($s=0$, $1$, and $2$).

\section{Beyond the Sinc approximation}
\label{sec_6}

\subsection{Spin-dependent corrections}
\label{sec_6_1}

In the high-frequency regime, the oscillatory structure of the absorption cross section for scalar fields is commonly described by the so-called sinc approximation, which captures the leading-order interference effect of surface waves trapped near the photon sphere~\cite{Decanini:2011xi}. However, this leading-order description neglects spin-dependent corrections, as well as higher-order modulations arising from the detailed structure of the Regge trajectories and the analytic form of the residues. These subleading effects are responsible for the so-called ``fine structure'' of the absorption spectrum, as discussed in detail by Décanini \textit{et al.}~\cite{Decanini:2011xw} in the case of massless scalar fields.

We now go beyond the sinc-approach to incorporate spin-dependent corrections into the oscillatory behavior of the absorption cross section. To this end, we insert the analytical expressions of the Regge poles and their associated residues, given respectively in Eqs.~\eqref{PR_WKB} and~\eqref{Residues_WKB}, into the Regge pole contribution Eq.~\eqref{Asymp_PR_cont}, restricting ourselves to the first Regge trajectory and its corresponding residue, which dominate the oscillatory component at high frequencies. To account for multiple circumnavigations of the surface wave around the photon sphere, we make use of the identity~\eqref{Rela_1} to rewrite the Regge pole series \( \sigma_{\text{RP}} \) in Eq.~\eqref{Asymp_PR_cont} as a harmonic sum. We retain only the leading harmonics \( m = 1 \) and \( m = 2 \), which capture the interference between the fundamental orbit and its higher harmonics. We then obtain the following expression for the Regge pole contribution \( \sigma_{s,\text{RP}}(\omega) \)
\begin{widetext}
\begin{equation}\label{Harmonic_Regge}
\begin{aligned}
\sigma_{s,\text{RP}}^{\text{(osc, harmonic)}}(\omega) =
\sigma_{\text{geo}}\bigg(&- 8\pi e^{-\pi} \frac{\sin\left[2\pi \, \Theta(\omega) + \delta_s(\omega) \right]}{2\pi \,\Theta(\omega)} + 16\pi e^{-2\pi} \frac{\sin\left[4\pi \, \Theta(\omega) + 2 \delta_s(\omega) \right]}{4\pi \,\Theta(\omega)}
\\
&-\frac{52\pi^2 e^{-\pi}}{9} \frac{\cos\left[2\pi \, \Theta(\omega) + \delta_s(\omega) \right]}{\left[2\pi \,\Theta(\omega)\right]^2} + \frac{208\pi^2 e^{-2\pi}}{9} \frac{\cos\left[4\pi \, 2\Theta(\omega) + \delta_s(\omega) \right]}{\left[2\pi \,\Theta(\omega)\right]^2}
\\
&+\frac{\Big(355 - 2930\pi + (3456\pi - 864)\beta \Big) \pi^3 e^{-\pi}}{243} \frac{\sin\left[2\pi \, \Theta(\omega) + \delta_s(\omega) \right]}{\left[2\pi \,\Theta(\omega)\right]^3} 
\\
&+\frac{\Big(355 - 5860\pi + (6912\pi - 864)\beta \Big) 8 \pi^3 e^{-2\pi}}{243} \frac{\sin\left[4\pi \, \Theta(\omega) + 2 \delta_s(\omega) \right]}{\left[4\pi \,\Theta(\omega)\right]^3} \bigg)
\end{aligned}
\end{equation}
\end{widetext}
where 
\begin{equation}\label{geo_phase}
\Theta(\omega) = 3\sqrt{3} M \omega,
\end{equation}
represents the ``geometric phase'' accumulated by surface waves orbiting near the photon sphere. It is directly related to the orbital period of a massless particle trapped on the photon sphere, given by \( T_\text{orb} = 2\pi b_c = 2\pi \times 3\sqrt{3} M \), where \( b_c \) is the critical impact parameter associated with null geodesics. The quantity \( \Theta(\omega) \) thus measures the number of wave cycles completed per revolution around the photon sphere. It governs the leading oscillatory behavior of the absorption cross section in the high-frequency regime. The interpretation of such oscillations in terms of trapped surface waves associated with Regge poles has been extensively developed in the literature~\cite{Decanini:2002ha,Decanini:2009mu,Decanini:2010fz,Andersson:1994rm}. The spin-dependent correction
\begin{equation}\label{shift_phase}
\delta_s(\omega) = \frac{\pi(65 - 72\beta)}{324\sqrt{3} M \omega}
\end{equation}
encodes subleading effects arising from diffraction and spin-curvature coupling in the effective potential. This structure is reminiscent of semiclassical interference phenomena, where a dominant geometric phase is modulated by a slowly varying spin-dependent shift, reflecting dispersive effects across the potential barrier.

The expression \eqref{Harmonic_Regge} provides a refined analytic representation of the oscillatory Regge pole contribution, incorporating not only the leading harmonic but also subleading modulations that account for the nonlinear structure of the Regge trajectory and the analytic form of the residues. However, in high frequencies regime, the dominant contribution to the oscillatory structure comes from the first harmonic term alone. By retaining only the leading-order sine term (corresponding to \( m = 1 \)) in the harmonic expansion and neglecting higher-order corrections, we obtain a simplified expression that still captures the essential physics of interference from surface waves orbiting near the photon sphere, while including the spin-dependent phase correction. This leads to the \textit{spin-dependent eikonal approximation}:
\begin{equation}\label{spin_eik_Regge}
\sigma_{s,\text{RP}}^{\text{(osc,eik)}}(\omega) = - 8\pi e^{-\pi}  \sigma_{\text{geo}} \frac{\sin\left[2\pi \, \Theta(\omega) + \delta_s(\omega) \right]}{2\pi \,\Theta(\omega)}
\end{equation}
This compact formula generalizes the sinc approximation introduced by Decanini \textit{et al.}~\cite{Decanini:2011xi} by incorporating a spin-dependent phase correction. As in the sinc approach, the prefactor \( 8\pi e^{-\pi} \) involves the Lyapunov exponent associated with the unstable photon sphere orbit followed by the particle~\cite{Decanini:2011xi,Decanini:2010fz,Cardoso:2008bp}. This formula provides an effective analytic tool for capturing the dominant oscillatory behavior of the absorption cross section for fields of various spins.

\begin{figure}[htbp]
 \includegraphics[scale=0.53]{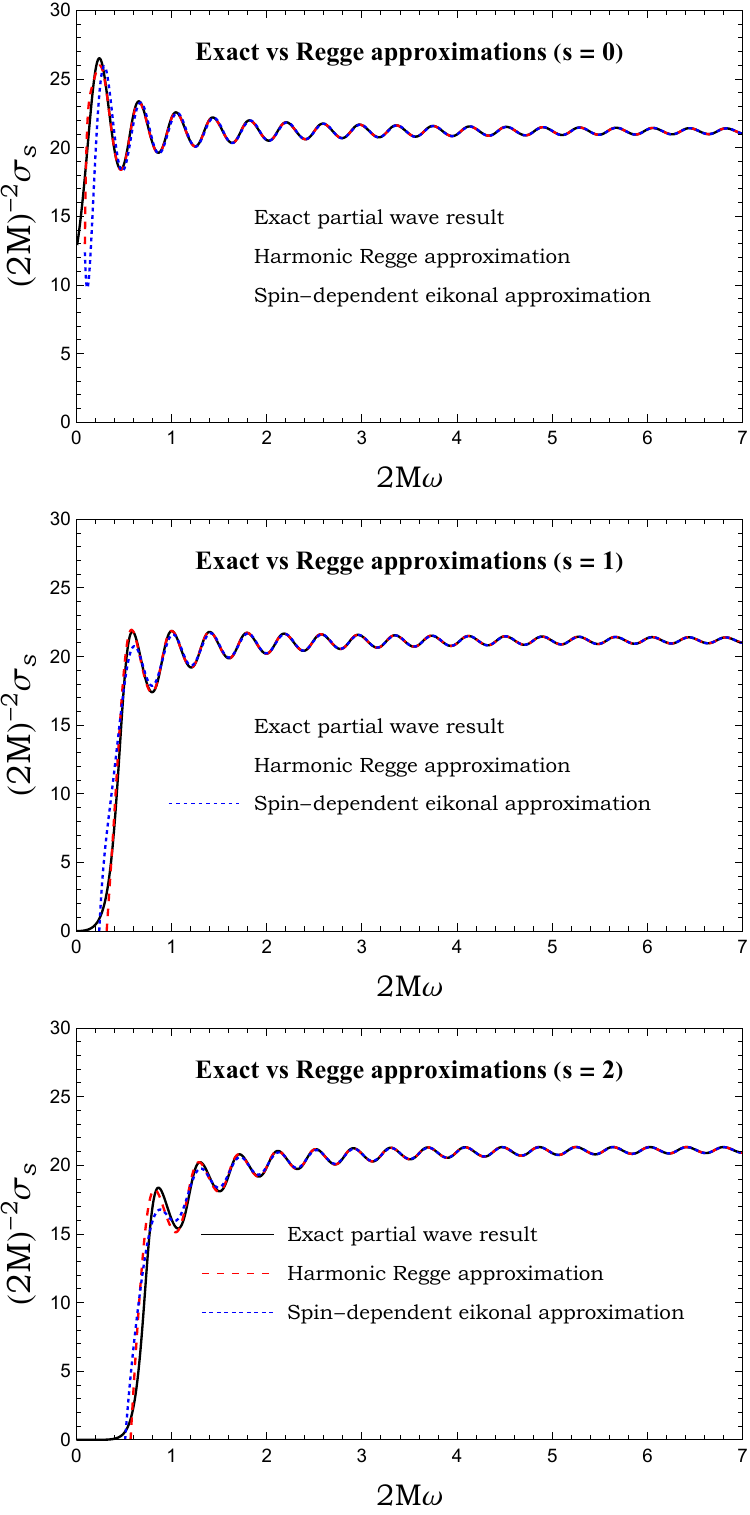}
\caption{\label{Fig:Spin-Sinc_Approx_Abs_s-0-1-2} Comparison between the total absorption cross section obtained from the partial wave expansion (black solid line) and two analytic CAM-based approximations for various field spins. Both approximations are constructed by inserting either the harmonic Regge contribution \( \sigma_{\text{RP}}^{\text{(osc, harmonic)}}(\omega) \) (red dashed line), given in Eq.~\eqref{Harmonic_Regge}, or the spin-dependent eikonal contribution \( \sigma_{\text{RP}}^{\text{(osc,eik)}}(\omega) \) (blue dotted line), given in Eq.~\eqref{spin_eik_Regge}, into the full CAM-based expression \( \sigma_s^{\text{approx}}(\omega) \), defined in Eq.~\eqref{Beyond_Sinc}. Top: scalar field (\( s = 0 \)); middle: electromagnetic field (\( s = 1 \)); bottom: gravitational field (\( s = 2 \)).}
\end{figure}
\begin{figure}[htbp]
 \includegraphics[scale=0.53]{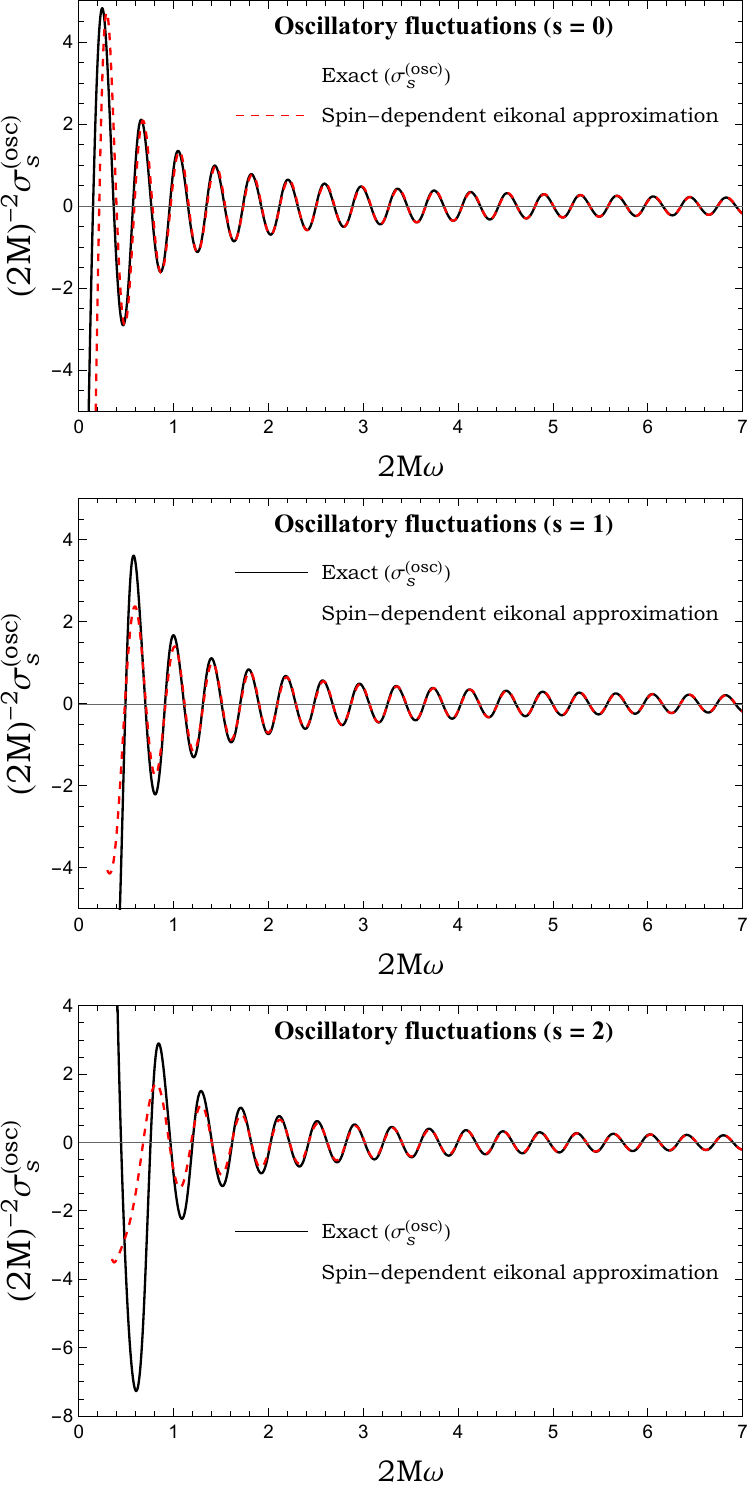}
\caption{\label{Fig:OSC_Spin-Eik_Approx_Abs_s-0-1-2} Oscillatory fluctuations in the total absorption cross section for different field spins. For each spin value, we display the difference between the exact total absorption cross section and the background absorption cross section $\sigma_s^{\text{osc}} = \sigma_s - (\widetilde{\sigma}_{s,\text{B}_\text{Re}} + \sigma_{s,\text{B}_\text{Im}} ) $ and compare it to the spin-dependent eikonal approximation (dashed red line) \eqref{spin_eik_Regge}. Top: scalar field (\( s = 0 \)); middle: electromagnetic field (\( s = 1 \)); bottom: gravitational field (\( s = 2 \)). This highlights the dominant oscillatory features associated with Regge pole interference, beyond the geometric background.}
\end{figure}

In Fig.~\ref{Fig:Spin-Sinc_Approx_Abs_s-0-1-2}, we compare the total absorption cross section computed from the partial wave expansion~\eqref{Sec_Absorption} with two analytic CAM-based approximations. The first, denoted \( \sigma_s^{\text{(approx, harmonic)}}(\omega) \), is constructed by inserting the harmonic Regge contribution \( \sigma_{\text{RP}}^{\text{(osc, harmonic)}}(\omega) \) from Eq.~\eqref{Harmonic_Regge} into the full CAM-based formula~\eqref{Beyond_Sinc}. The second, \( \sigma_s^{\text{(approx, eik)}}(\omega) \), includes only the leading spin-dependent eikonal contribution \( \sigma_{\text{RP}}^{\text{(osc, eik)}}(\omega) \), given in Eq.~\eqref{spin_eik_Regge}. As shown in the figure, both approximations agree remarkably well with the exact result in the high-frequency regime (\( 2M\omega \gtrsim 1 \)). However, as the frequency decreases into the intermediate and low-frequency ranges, the harmonic Regge approximation offers a visibly improved match, especially for electromagnetic and gravitational fields. Interestingly, for scalar fields (\( s = 0 \)), the spin-dependent eikonal approximation remains accurate down to lower frequencies.

An important subtlety arises when comparing the effectiveness of Regge-based approximations across different spin fields. In particular, the inclusion of the subleading \( \mathcal{O}(1/\omega) \) correction in the Regge pole phase proves crucial for accurately capturing the absorption spectrum of electromagnetic $(s = 1)$ and gravitational $(s=2)$ fields in the intermediate-frequency regime (see Fig.~\ref{Fig:Spin-Sinc_Approx_Abs_s-0-1-2}). This correction encodes nongeometric effects arising from diffraction and spin-dependent potential curvature near the photon sphere, which are especially relevant before the wave dynamics becomes fully governed by null geodesics. This observation can be traced back to the structure of the QNM spectrum: for electromagnetic and gravitational perturbations, the real part of the lowest QNM appears only beyond \( 2M\omega_{\ell =1, n=1} \sim 0.496 \) and \( 2M\omega_{\ell=2, n=1}\sim 0.747 \), respectively. Below these frequencies, the surface waves are not yet tightly localized around the photon sphere, and the phase accumulation deviates significantly from the eikonal limit. In this context, the leading-order geometric phase, such as the one appearing in the sinc approximation proposed by Decanini \textit{et al.}, fails to capture the correct interference structure unless supplemented by subleading corrections. In contrast, for scalar field (\( s = 0 \)), the first QNM emerges at much lower frequencies (\( 2M\omega_{\ell=0,n=1} \sim 0.221 \)), allowing the surface waves to quickly settle into a quasigeodesic regime. As a result, the sinc approximation, despite neglecting the \( \mathcal{O}(1/\omega) \) correction, remains remarkably accurate over a wider frequency range. This reinforces the interpretation of the \( 1/\omega \) term in the Regge phase as a diagnostic of the transition from diffractive to geometric-optics behavior, whose onset is intrinsically spin-dependent.

To conclude this subsection, we display in Fig.~\ref{Fig:OSC_Spin-Eik_Approx_Abs_s-0-1-2} the oscillatory fluctuations of the total absorption cross section for different field spins. For each spin value, we plot the difference between the exact total absorption cross section and the background contribution, defined as 
\begin{equation}
\sigma_s^{\text{osc}}(\omega)  = \sigma_s(\omega) - \left[\widetilde{\sigma}_{s,\text{B}_\text{Re}}(\omega) + \sigma_{s,\text{B}_\text{Im}}(\omega) \right], 
\end{equation}
and compare it with the spin-dependent eikonal approximation given in Eq.~\eqref{spin_eik_Regge}. The agreement is remarkable, showing that the oscillatory structure of the total absorption cross section around the background absorption cross section is already very well captured by the contribution of the first Regge pole across all field spins considered in the Schwarzschild BH.

\subsection{Fine structure of high-energy absorption cross sections}
\label{sec_6_2}

\begin{figure}[htbp]
 \includegraphics[scale=0.53]{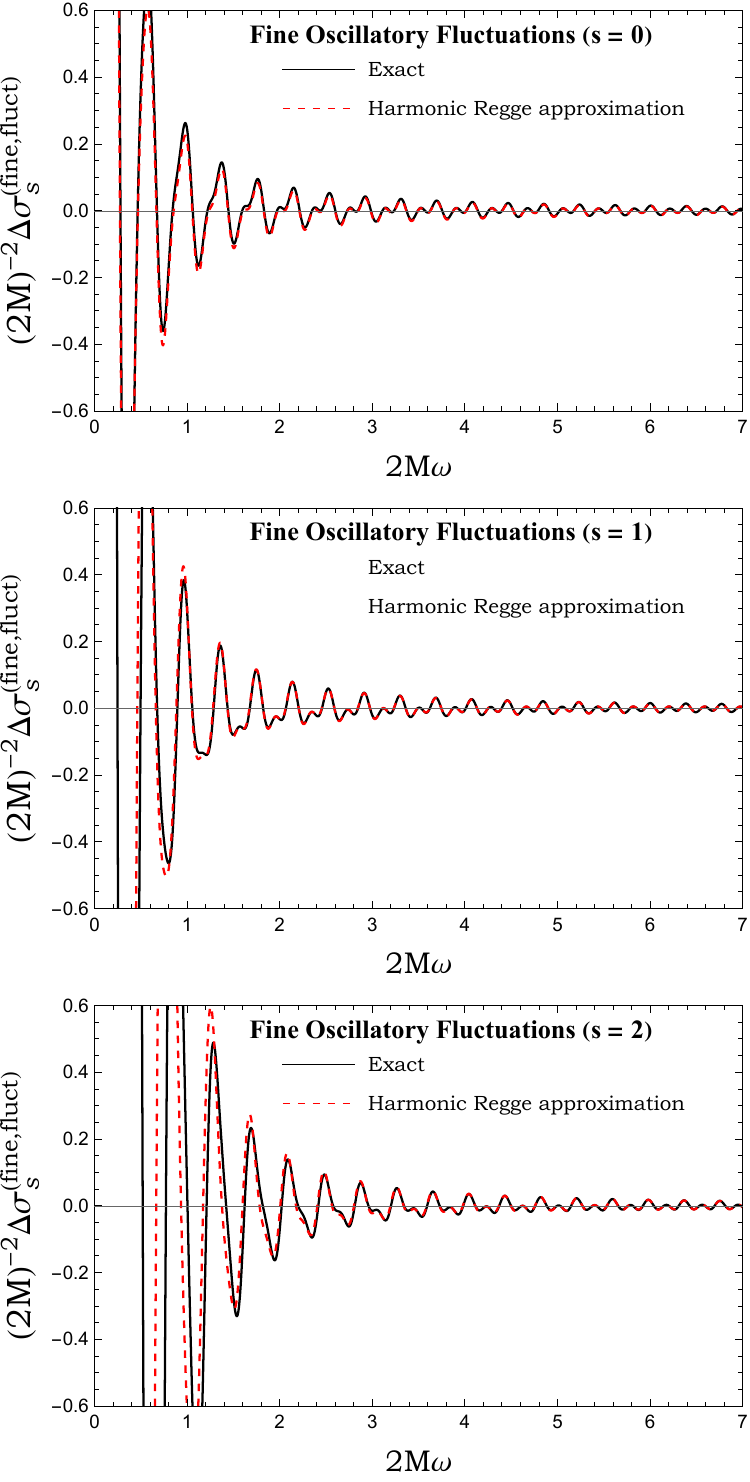}
\caption{\label{Fig:fluct_Abs_s-0-1-2} Residual fine structure in the absorption cross section beyond the spin-dependent eikonal approximation. The black curve shows the difference \( \Delta\sigma_s^{\text{(fine,fluct)}} = \sigma_s - \sigma_s^{\text{(approx, eik)}} \), obtained by subtracting the spin-dependent eikonal approximation from the exact cross section. The red dashed curve shows the difference between the harmonic and eikonal CAM-based approximations, \( \sigma_s^{\text{(approx, harmonic)}} - \sigma_s^{\text{(approx, eik)}} \), demonstrating that the harmonic approximation accurately captures the subleading oscillatory modulations. Top: scalar field (\(s = 0\)); middle: electromagnetic field (\(s = 1\)); bottom: gravitational field (\(s = 2\)).}
\end{figure}

Although the agreement between the exact absorption cross section and the spin-dependent eikonal approximation may appear nearly perfect to the naked eye in the high-frequency regime, a more refined analysis reveals residual oscillatory features beyond the leading behavior. To isolate this fine structure, we consider the difference between the exact cross section and the spin-dependent eikonal approximation
\begin{equation}\label{Fine_fluct}
  \Delta\sigma_s^{\text{(fine,fluct)}}(\omega) = \sigma_s(\omega) - \sigma_s^{\text{(approx, eik)}}(\omega).
\end{equation}
As shown in Fig.~\ref{Fig:fluct_Abs_s-0-1-2}, this residual clearly displays a fine structure in the absorption cross section across all spin values. The amplitude of this structure lies in the range of approximately \( 5\%\text{--}10\% \) for scalar and electromagnetic fields (\( s = 0,1 \)), and can reach up to \( 21\% \) in the gravitational case (\( s = 2 \)).  To verify that this fine structure is indeed captured by the harmonic expansion, we directly compare the two CAM-based approximations
\begin{equation}
\sigma_s^{\text{(approx, harmonic)}}(\omega) - \sigma_s^{\text{(approx, eik)}}(\omega).
\end{equation}
The resulting difference reproduces the fine structure with excellent fidelity, confirming that the inclusion of higher harmonics in the Regge pole series accurately captures the subleading oscillatory features that are absent in the leading-order eikonal description.

It is important to note that for the scalar case (\( s = 0 \)), this fine structure has already been studied in detail by Décanini \textit{et al.}~\cite{Decanini:2011xw}. In particular, by neglecting the Regge phase correction \( \delta_s(\omega) \) and truncating the expansion at \( \mathcal{O}(1/(2M\omega)^2) \), one recovers their analytical expression given in Eq.~(2.12) of~\cite{Decanini:2011xw}. As emphasized in~\cite{Decanini:2011xw}, the fine structure arises entirely from the surface waves trapped near the photon sphere and is fully encoded in the contribution of the first Regge pole. Physically, this phenomenon results from interference between partial waves completing multiple orbits around the BH, with the dominant oscillation governed by the orbital period of a massless particle on the photon sphere. Subleading modulations, often referred to as ``beats,'' emerge from interference between the fundamental mode and its higher harmonics. Moreover, the amplitude of these oscillations reflects the dispersive character of the Regge trajectory: higher-order corrections to the Regge pole position modulate both the phase and amplitude of the signal, thereby encoding spin-curvature interactions and diffraction effects beyond the leading geometric optics limit.

\section{Conclusion}
\label{sec_7}

The study of the absorption cross section of a Schwarzschild black hole for massless bosonic fields with spins \( s = 0, 1, 2 \) reveals a rich structure, especially when analyzed through the formalism of complex angular momentum. This approach highlights the subtle influence of spin on the absorption process, both in the low- and high-frequency regimes.

We have constructed a complete CAM-based representation of the absorption cross section, expressed as the sum of background contributions arising from integrals along the real and imaginary axes in the CAM-plane, together with a discrete Regge pole series. This decomposition allowed us to identify three distinct physical mechanisms contributing to the absorption process: ``classical geometric'' propagation associated with the real-axis contribution, subleading corrections from the imaginary-axis contribution, and interference effects associated with surface waves orbiting near the photon sphere.

Going beyond the geometric optics description and the sinc-type approximation originally derived for scalar fields, we have first developed an improved analytic formula that captures both the dominant oscillatory modulations and the fine structure of the absorption spectra for massless fields of spin $s = 0$, $1$, and $2$. This expression incorporates spin-dependent phase corrections and higher-order effects associated with surface wave interference. Subsequently, by neglecting subleading corrections, we have derived a simplified formula that generalizes the sinc approximation, accurately describing the dominant oscillations for spin-$s=1$ and $s=2$ fields. Together, these results provide a comprehensive and unified semiclassical description of black hole absorption across frequency regimes.

It is also important to emphasize the duality between Regge poles and the complex frequencies of weakly damped QNMs of black holes~ \cite{Decanini:2002ha,Decanini:2009mu,Decanini:2010fz}. This correspondence offers a compelling interpretation of both the high-frequency fluctuations and the fine structures of the absorption cross section in terms of QNM physics.

Our framework provides a unified and physically consistent interpretation of the spin-dependent features of black hole absorption. It establishes a solid foundation for further generalizations, including extensions to more complex geometries (e.g., rotating or charged black holes) and for observational applications such as black hole shadows, strong gravitational lensing, or even Hawking radiation. Several of these directions are currently under investigation.

\section*{Acknowledgments}

We are grateful to Sam R. Dolan for his insightful suggestions and constructive feedback, which have significantly improved this work. We also thank the referee for their valuable comments and suggestions, which helped clarify key points and enrich the paper.

\bibliography{Absorp_Cross_Sec_Spins}

\end{document}